\documentclass[twocolumn,preprintnumbers,amsmath,amssymb,superscriptaddress,longbibliography,nofootinbib]{revtex4-1}
\usepackage{filecontents}
\usepackage{graphicx}
\usepackage{bm}
\usepackage{dsfont}
\usepackage[usenames,dvipsnames]{xcolor}
\usepackage[tight]{subfigure}
\usepackage{verbatim}
\usepackage{units}
\usepackage{multirow}
\usepackage{enumitem}
\usepackage{mathrsfs}
\usepackage{leftidx}
\usepackage{xspace}
\usepackage{braket}
\usepackage{bbm}
\usepackage{cancel}
\usepackage{amsfonts,amssymb,amsmath}
\usepackage[normalem]{ulem}

\usepackage{xr}

\usepackage[a4paper]{hyperref}
\hypersetup{colorlinks=true,linktoc=all,linkcolor=blue,breaklinks=true,citecolor=blue,urlcolor=blue}
\usepackage[english]{babel}
\usepackage{cleveref}

\newcommand{\kB}{k_\text{B}}

\newcommand{\tr}[1]{\text{tr}\left\{ {#1} \right\}}

\newcommand{\Tr}{{\rm tr}}
\newcommand{\be}{\begin{equation}\small\begin{aligned}}{}
\newcommand{\ee}{\end{aligned}\end{equation}}

\newcommand{\Var}{\mathrm{Var}}
\newcommand{\MRH}{\mathrm{H}}
\newcommand{\MRL}{\mathrm{L}}

\newcommand{\MRS}{\mathrm{S}}
\newcommand{\MRR}{\mathrm{R}}

\newcommand{\SKUR}{S-KUR}
\newcommand{\curlyJ}{ \mathcal{J}^{(X)}_\alpha}
\newcommand{\Mdot}{\dot{M}^{(X)}(t)}

\begin{document}
\title{Susceptibility-kinetic uncertainty relations for quantum systems}

\author{Didrik Palmqvist}
	\affiliation{Department of Microtechnology and Nanoscience (MC2), Chalmers University of Technology, S-412 96 G\"oteborg, Sweden}

\author{Janine Splettstoesser}
	\affiliation{Department of Microtechnology and Nanoscience (MC2), Chalmers University of Technology, S-412 96 G\"oteborg, Sweden}

\author{Ludovico Tesser}
	\affiliation{Department of Microtechnology and Nanoscience (MC2), Chalmers University of Technology, S-412 96 G\"oteborg, Sweden}

\date{\today}

\begin{abstract}
Kinetic uncertainty relations bound current precision of stochastic processes by dynamical activity. The extension of these bounds to quantum systems has been impeded by coherence, strong system-reservoir coupling, and the subtlety of defining dynamical activity in the quantum regime. Here, we introduce a \emph{partial dynamical activity} through the quantum Fisher information associated with the rescaling of the system-reservoir coupling and show that it bounds current precision via a universal \emph{susceptibility-kinetic uncertainty relation}. The general validity of this relation is guaranteed by the natural contribution of a susceptibility term, which is experimentally accessible by tuning the system-reservoir coupling strength. 
We show how the partial dynamical activity encompasses previous definitions of activity in the weak-coupling Markovian limit and that it provides an information-geometric interpretation of correlator-based activities.
We illustrate the tight constraint on precision that our bound provides with the example of the dynamics of a strongly interacting finite-size system and of steady-state coherent transport through a double quantum dot, where quantum effects invalidate previously developed kinetic uncertainty relations. We expect our bound to provide a powerful tool for optimizing precision in arbitrary quantum systems. 
\end{abstract}

\maketitle

\section{Introduction}

\begin{figure}[b!]
    \centering
   \includegraphics[width=2.7in]{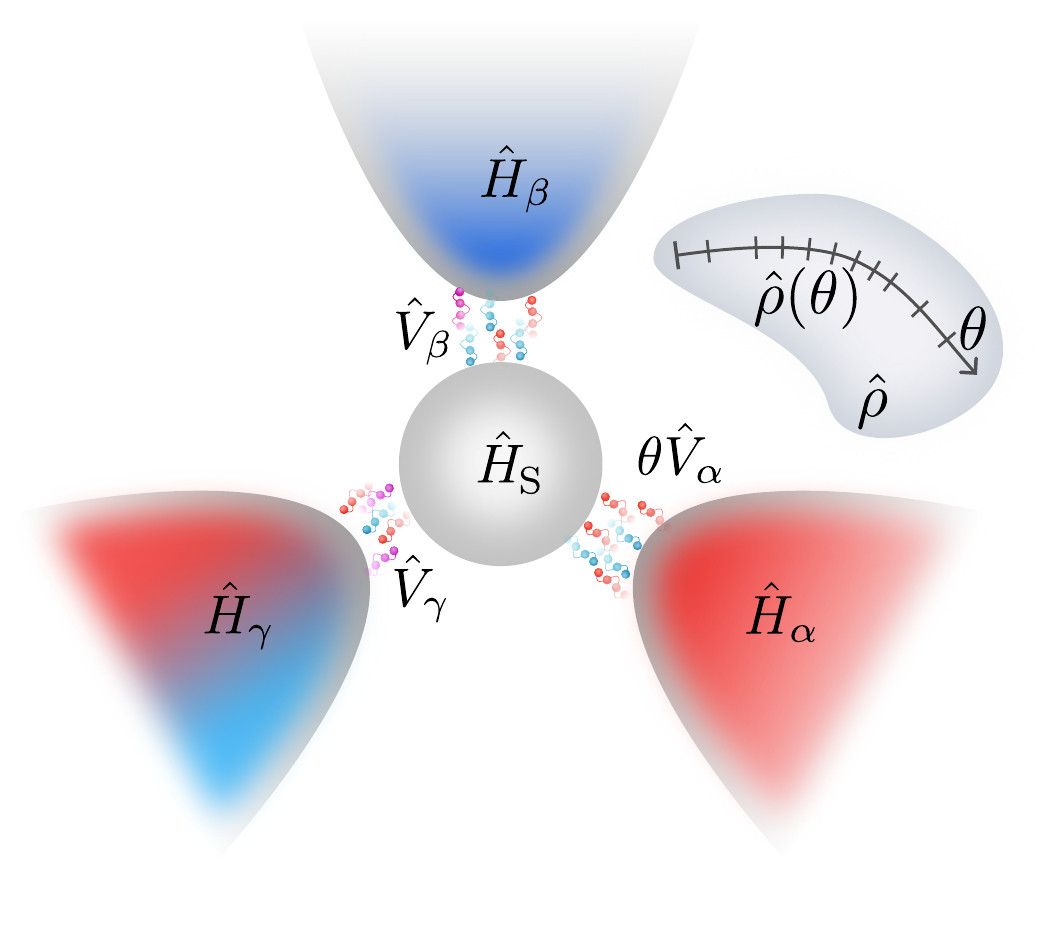}
\caption{Quantum system S coupled to reservoirs via coupling $\hat{V}_{\alpha,\beta,...}$, enabling transport. The coupling strength to (one) reservoir is parameterized by $\theta$, resulting in its encoding in the state dynamics $\hat{\rho}(\theta)$ that serves as a basis for generalizing the concept of dynamical activity to quantum systems. The color-shading of the reservoirs symbolizes arbitrary out-of-equilibrium situations.}
    \label{fig: sketch general system}
\end{figure}

Fluctuations play a central role in small-scale systems: They carry information about microscopic dynamics and directly affect precision and performance of devices ranging from sensors to thermodynamic machines such as heat engines and refrigerators.
These tasks are typically performed \textit{out of equilibrium}, where celebrated equilibrium relations such as the fluctuation-dissipation theorem~\cite{Johnson1927Jan, Nyquist1928Jul} are no longer universally applicable, outside of specific exceptions~\cite{Altaner2016Oct, Dechant2020Mar}.
Establishing universal relations constraining fluctuations out of equilibrium is challenging, but provides fundamental insights into the dynamics and thermodynamics of small-scale systems.
In recent years, important progress in establishing fluctuation bounds has been made through the discovery of the thermodynamic and kinetic uncertainty relations. Originally derived for classical Markovian dynamics, the thermodynamic uncertainty relations (TURs) limit precision by entropy production~\cite{Barato2015Apr,Gingrich2016Mar, Horowitz2020Jan,Koyuk2020Dec} while the kinetic uncertainty relations (KURs) bound precision by the dynamical activity, i.e. the number of jumps observed in a stochastic trajectory~\cite{Garrahan2017Mar,DiTerlizzi2018Dec, Yan2019Sep, Hiura2021May,Liu2025Feb}.
These aspects have been extended to more general relations~\cite{Aslyamov2025Apr} and unified in the thermokinetic uncertainty relation (TKUR)~\cite{Vo2022Sep}. Such uncertainty relations constitute performance limits, quantifying the cost of achieving high precision, but can also be exploited as inference tools, where difficult-to-measure quantities, such as entropy production and dynamical activity, can be inferred from current and noise measurements~\cite{Seifert2019Mar}.

In quantum systems these fluctuation bounds can be broken, indicating precision advantages induced by quantum effects~\cite{brandner_thermodynamic_2018,Agarwalla2018Oct,Saryal2019Oct,potanina_thermodynamic_2021,Prech2023Jun,Manzano2023Oct,Hasegawa2020Jul,Hasegawa2023May,Prech2025Jan,VanVu2025Mar,Meier2025Jul,VanVu2026May,Brandner2025Jul,Brandner2025Oct,Hasegawa2025Nov,BlasiMay2025}. It is therefore of crucial importance to expand the validity of uncertainty relations to the quantum regime. Previous attempts to extend precision bounds to the quantum regime have focused on weak system-reservoir coupling~\cite{Carollo2019Apr,Hasegawa2020Jul,VanVu2022Apr,Hasegawa2023May,Prech2025Jan,VanVu2025Mar,Liu2026Jun}, weak particle-particle interactions or weak nonlinearities~\cite{brandner_thermodynamic_2018,potanina_thermodynamic_2021,Palmqvist2025Oct,Brandner2025Jul,Palmqvist2025Jul,BlasiMay2025, Tesser2026Jul}, or have exploited measurement-based methods, which are typically less predictive in the long-time limit~\cite{Hasegawa2025Nov,VanVu2026May}. A further drawback of these quantum bounds is that they are often restricted to the precision of specific transport quantities or that quantum corrections are cumbersome and not directly accessible experimentally. This is particularly striking for quantum extensions of KURs, since the concept of ``activity" in quantum systems is not even uniquely defined.
Predictive and meaningful precision bounds, valid for strongly-coupled, interacting systems in the long-time limit, have hence been lacking.

In this work, we establish a general kinetic uncertainty relation, which overcomes these restrictions. We accomplish this by introducing a partial dynamical activity (PDA) via the quantum Fisher information, quantifying exchanges of a central system with an environment (or parts of the environment), see Fig.~\ref{fig: sketch general system}, and show that it bounds the precision of generic observables, 
\begin{equation}\label{eq: intro ss transport MKUR}
  \begin{aligned}
       \frac{\left( \int_0^t ds \mathcal{J}^{(X)}(s)  \right)^2}{\text{Var}[\Delta \hat{X}^\MRH(t)]} 
   &\leq \mathscr{A}_V(t) .\\
  \end{aligned}
\end{equation} 
This bound, derived employing the quantum Cram\'er-Rao bound~\cite{Helstrom1969Jun}, is the central result of this paper.
The partial dynamical activity $\mathscr{A}_V(t)$ constrains the average of the change of the observable $\Delta\hat{X}^\MRH(t) = \hat{X}^\MRH(t)-\hat{X}^\MRH(0)$ and its variance $\text{Var}[\Delta \hat{X}^\MRH(t)]$ via the integrated response current $\int_0^t ds \mathcal{J}^{(X)}(s) =\frac{1}{2} (\langle \Delta\hat{X}^\MRH(t) \rangle +M^{(X)}(t))$ using the \textit{susceptibility} $M^{(X)}(t)$. This susceptibility, introduced in detail in Sec.~\ref{sec:MKUR}, captures the response of the observable to a coupling strength.
Importantly, it guarantees the general validity of the KUR in the presence of quantum effects, and it is measurable whenever this coupling strength is tunable in an experiment.
We therefore refer to the bound~\eqref{eq: intro ss transport MKUR} as the \textit{susceptibility-kinetic uncertainty relation} (\SKUR). In the incoherent weak-coupling limit, we have $\int^t_0 ds\mathcal J_\alpha^{(X)}(s) \to \langle \Delta\hat{X}^\MRH(t)\rangle$ and we recover a classical KUR.
We collect an overview of all key quantities in Table~\ref{tab:overview}.

In addition, the \SKUR\ is equivalently reformulated as precision bound 
\begin{equation}\label{eq: intro ss precision SKUR}
  \begin{aligned}
       \frac{ \langle \Delta\hat{X}^\MRH(t)\rangle^2}{\text{Var}[\Delta \hat{X}^\MRH(t)]} 
   &\leq \frac{4\mathscr{A}_V(t)}{ (1 + \Phi^{(X)}(t))^2} ,\\
  \end{aligned}
\end{equation} 
where $\Phi^{(X)}(t) = M^{(X)}(t)/  \langle \Delta\hat{X}^\MRH(t)\rangle $ acts as a quantum correction, with a structure resembling the weak-coupling quantum corrections of Refs.~\cite{Prech2025Jan,Ptaszynski2019Nov}.

The \SKUR\ is universal in the sense that it does not rely on weak-coupling assumptions, Markovian dynamics, noninteracting particles, or on taking the steady-state limit, but holds even for transient regimes and time-dependent driving.
The \SKUR\ is particularly well suited to constraining precision in quantum transport, since change of a reservoir observable $\hat{X}_\alpha$, such as charge, particle number or energy is typically induced by the coupling between system and environment $\hat{V}_\alpha(t)$, for which a \emph{local} activity, with respect to this coupling enters the \SKUR~\eqref{eq: intro ss transport MKUR}. 

We show that our \SKUR~\eqref{eq: intro ss transport MKUR} recovers the classical limit in regimes where quantum effects can be neglected. We furthermore show that the PDA is limited by the correlator-based activity proposed in Ref.~\cite{BlasiMay2025}, and is connected to the notions of activity obtained with time rescaling~\cite{Hasegawa2023May,Nakajima2023Nov, Nishiyama2024Apr, Nishiyama2025Apr} as well as with measurement probability~\cite{Hasegawa2025Nov,VanVu2026May}. Our main result~\eqref{eq: intro ss transport MKUR} hence offers an overarching, generally valid bound implying and improving previously introduced limiting cases and we expect it to provide important guidelines for the design of future high-precision quantum devices.

We apply the \SKUR\ to the analysis of the precision of charge- and energy-transfer in two complementary double quantum-dot setups: steady-state coherent transport through a noninteracting double dot, as well as dynamic particle transfer through a strongly interacting double dot with finite-size reservoirs. Double quantum dots are highly relevant for controlled and tunable electron transport, for the implementation of thermal machines, and for sensing schemes~\cite{vanderWiel2002Dec,Barthel2010Apr}. A consistent analysis of their potential for increased precision due to quantum effects is hence required. At the same time, previously established precision bounds break down for transport through double dots, since the coherent superpositions between quantum-dot states lead to nontrivial quantum effects when the system is brought out of equilibrium. We show that even in regimes where standard KURs are strongly violated and where the precision of the device is potentially high, the susceptibility contribution to $\curlyJ$ guarantees the validity of the \SKUR~\eqref{eq: intro ss transport MKUR}. Furthermore, we compare the predictiveness of the \SKUR\ with the recently proposed bounds for steady-state transport of Refs.~\cite{Brandner2025Jul,BlasiMay2025} for the DQD.

The remainder of the paper is organized as follows: We derive the \SKUR\ in Sec.~\ref{sec:MKUR} and show its specific implementation for a generic quantum-transport situation. We then show how the \SKUR\ and its constraint given by the PDA connect to previously established results in the weak-coupling limit, Sec.~\ref{sec:relate_weak coupling}, for definitions based on information geometry, Sec.~\ref{sec:relate_total}, and for measurement-based approaches, Sec.~\ref{sec:relate_measurement based}. Next, we demonstrate the predictiveness and validity of the \SKUR\ at the experimentally and technologically relevant example of quantum transport through a double quantum dot in Sec.~\ref{sec:dqd}. Finally, we apply the \SKUR\ in an interacting DQD coupled to finite-size leads in Sec.~\ref{sec: DQD finite leads}. Technical details of derivations and limiting cases are presented in the appendices together with complementary plots.

\section{Susceptibility-KUR}\label{sec:MKUR}

Here, we derive the \SKUR~\eqref{eq: intro ss transport MKUR} starting from a formulation for a general quantum system, where the current of an observable $\hat{X}$ is induced by a part  $\hat{V}(t)$ of the total Hamiltonian. In order to bound the precision of this current, we use the quantum Cram\'er-Rao bound for the response of the observable to the strength of $\hat{V}(t)$, parametrized by $\theta$.
We identify the corresponding quantum Fisher information as the \textit{partial dynamical activity} (PDA) quantifying the dynamics generated by $\hat{V}(t)$. This PDA is shown to bound the precision of the observable response in the \SKUR.

The \SKUR\ is not limited to but can naturally be applied to generic open quantum systems and transport settings, where $\hat{V}(t)$ describes the coupling between a system and (a part of) its environment. In these cases  the \SKUR\ constrains the precision in the transport of typical reservoir observables, such as, e.g., charge and energy.

\subsection{\SKUR\ for generic quantum systems}
In this section, we prove the \SKUR, Eq.~\eqref{eq: intro ss transport MKUR}, for a general quantum system. We consider an observable $\hat{X}$, which, in Heisenberg picture, evolves according to 
\begin{equation}\label{eq:current}
\begin{split}  
    \hat{J}^{(X)}(t):= \partial_t\hat{X}^\MRH(t) =& \frac{i}{\hbar}[\hat{H}^\MRH(t ), \hat{X}^\MRH(t)] \\
    &= \frac{i}{\hbar}[ \hat{V}^\MRH(t), \hat{X}^\MRH(t)],
\end{split}
\end{equation}
with an associated current operator $\hat{J}^{(X)}(t)$.
This evolution of the operator yields a natural split of the Hamiltonian $\hat{H}(t) = \hat{H}_0(t) +\hat{V}(t)$ into the generator of the dynamics of $\hat{X}^\MRH(t)$, namely $\hat{V}(t)$, and the remaining term, $\hat{H}_0(t)$, which could in principle even be zero.
We will later see the example of transport where $\hat{X}$ is a reservoir's charge or energy, and $\hat{V}(t)$ is the coupling Hamiltonian to a given reservoir, where currents are detected.

Assuming that the initial state $\hat{\rho}(0)$ commutes with $\hat{X}$, $[\hat{\rho}(0),\hat{X}] =0$, the change in time of the operator
\begin{equation}\label{eq:DeltaX}
    \Delta\hat{X}^\MRH(t)  = \hat{X}^\MRH(t) -\hat{X}
\end{equation}
has appealing statistical properties.
Specifically, the mean and the variance of the change of the operator,
\begin{equation}
    \begin{aligned}
        \langle \Delta \hat{X}^\MRH(t)  \rangle &:= \tr{\Delta \hat{X}^\MRH(t) \hat{\rho}(0)},\\ \mathrm{Var}[\Delta \hat{X}^\MRH(t) ] &:=\tr{ (\Delta \hat{X}^\MRH(t) )^2 \hat{\rho}(0)}-\langle \Delta \hat{X}^\MRH(t)  \rangle ^2,
    \end{aligned}
\end{equation}
coincide with those of the two-point measurement (TPM) scheme\footnote{{Higher moments of $\Delta \hat{X}^\MRH(t)$ need not coincide with the TPM. Precision bounds as the one presented here involve only the first two moments.}}~\cite{Esposito2009Dec}.

 We use the quantum Cram\'er-Rao bound~\cite{Helstrom1969Jun} to constrain the precision of the response of the observable $\Delta \hat{X}^\MRH(t)$ to a chosen parameter.
Since $\hat{V}(t)$ induces a change in $\hat{X}$, we parametrize its strength,
\begin{equation}\label{eq:Htheta}
    \hat{H}_\theta(t) = \hat{H}_0(t) + \theta \hat{V}(t).
\end{equation}
Note that we do \textit{not} require $\hat{V}(t)$ to be weak. To use the Cram\'er-Rao bound, we imprint the parametrization on the evolution of the system state via 
\begin{equation}
    \hat{U}_\theta(t,0) := T_+ \exp\left\{-\frac{i}{\hbar} \int^t_0 ds \hat{H}_\theta(s) \right\},
\end{equation}
such that the state at time $t$ is given by
\begin{equation}
    \hat{\rho}(t,\theta) = \hat{U}_\theta(t,0) \hat{\rho}(0) \hat{U}^\dag_\theta(t,0).
\end{equation}
Since in the Cram\'er-Rao bound all the dependence on $\theta$ is carried by the state $\hat{\rho}(t, \theta)$, we introduce the Schr\"odinger picture operator
\begin{equation}
\hat{Y} := \hat{X} - \hat{U}(t,0) \hat{X} \hat{U}^\dag(t,0), 
\end{equation}
with $\hat{U}(t,0) = \hat{U}_{\theta=1}(t,0)$.
This operator fulfills $\left.\tr{ \hat{Y} \hat{\rho}(t,\theta) }\right|_{\theta=1} = \langle \Delta \hat{X}^\MRH(t) \rangle$ and, under the natural assumption $[\hat{\rho}(0), \hat{X}]=0$, also the response of the average $\Delta \hat{X}^\MRH$ and $\hat{Y}$ coincide, 
\begin{equation}
    \partial_\theta \langle \Delta \hat{X}^{\MRH}(t,\theta) \rangle|_{\theta=1} =  \partial_\theta \tr{ \hat{Y} \hat{\rho}(t,\theta) }\Big|_{\theta=1},
\end{equation}
see also Appendix~\ref{app:Quantum-Fisher-Information}. With this, the quantum Cram\'er-Rao bound~\cite{Helstrom1969Jun} yields
\begin{equation}\label{eq:CR-KUR}
    \frac{\left(\partial_\theta \langle \Delta \hat{X}^\MRH(t,\theta) \rangle \big|_{\theta=1} \right)^2}{\Var[\Delta \hat{X}^\MRH(t) ]} \leq \left.\mathcal{F}_{\theta\theta}\right|_{\theta=1}=:4 \mathscr{A}_V(t),
\end{equation}
where $\mathcal{F}_{\theta\theta}$ is the quantum Fisher information (QFI). The QFI is a central quantity in quantum metrology and information geometry~\cite{Liu2019Dec}, see Appendix~\ref{app:Quantum-Fisher-Information} for its general definition.
We here introduce the  concept of a \textit{partial dynamical activity} (PDA) $\mathscr{A}_V(t)=(\left.\mathcal{F}_{\theta\theta}\right|_{\theta=1})/4$ with respect to the coupling $\hat{V}(t)$ which corresponds to the Bures metric~\cite{Liu2019Dec}. The PDA is ``partial" in the sense that it is chosen to only account for the activity due to the part of the dynamics governing $\hat{X}^\MRH(t)$. In the following, we (i) justify that the quantity $\mathscr{A}_V(t)$ is indeed a measure of activity and (ii) show how the current precision of interest enters the Cram\'er-Rao bound~\eqref{eq:CR-KUR}, thereby establishing a quantum KUR.

Evaluating the PDA, we find
\begin{equation}\label{eq:A_V}
\begin{split}
    \mathscr{A}_V(t) &=\sum_{ij} \frac12\frac{(p_i-p_j)^2}{p_i+p_j}|\braket{i|\int_0^t\frac{ds}{\hbar}\hat{V}^\MRH (s)|j}|^2,
\end{split}
\end{equation}
where $p_i$ and $\ket{i}$ are the eigenvalues and eigenstates of the initial total system-environment state $\hat{\rho}(0) = \sum_i p_i\ket{i}\!\!\bra{i}$.
Since the terms with $i=j$ do not contribute to the sum in Eq.~\eqref{eq:A_V}, we can replace $\hat{V}^\MRH(s)\to \delta\hat{V}^\MRH(s):= \hat{V}^\MRH(s) - \langle \hat{V}^\MRH(s)\rangle$. Furthermore, using $[(p_i-p_j)/(p_i+p_j)]^2\leq 1$, we find that the partial dynamical activity of Eq.~\eqref{eq:A_V} is limited by the fluctuations in the coupling Hamiltonian
\begin{equation}\label{eq:Alim}
    \mathscr{A}_V(t)\leq \!\!\int_0^t \!\!\frac{ds}{\hbar}\!\!\int_0^t \!\!\frac{ds'}{\hbar}\langle \delta\hat{V}^\MRH(s)\delta\hat{V}^\MRH(s')\rangle=: \mathscr{A}_V^\text{lim}(t).
\end{equation}
The fluctuations in the coupling Hamiltonian $\mathscr{A}_V^\text{lim}(t)$ are often easier to calculate than the partial dynamical activity $\mathscr{A}_V(t)$, and the two coincide for pure initial states, where only one $p_i$ is equal to one, and the rest is zero. This relation between $\mathscr{A}_V(t)$ and $\mathscr{A}_V^\text{lim}(t)$ is explained by the fact that the PDA contains only the correlations in $\hat{V}^\MRH(t)$ related to how sensitive the state is to the coupling strength, while the limiting activity also includes fluctuations stemming from the uncertainty in the initial ensemble. 

Even if the initial state is mixed, $\mathscr{A}_V^\mathrm{lim} (t)$ possesses a rigorous information-geometric interpretation.
Given the spectral decomposition of the physical system-environment initial state $\hat{\rho}(0) = \sum_i p_i\ket{i}\!\!\bra{i}$, we introduce a purification using a fixed reference system R~\cite{Nielsen2010Dec,Watrous2018Apr}
\begin{equation}\label{eq:purified_state}
    \ket{\Psi(0)} = \sum_i \sqrt{p_i} \ket{i} \ket{\mathrm{R}_i},
\end{equation}
where the reference states are orthonormal $\braket{\mathrm{R}_i|\mathrm{R}_j} = \delta_{ij}$. Only the physical system evolves in time while the reference is kept fixed:
\begin{equation}
    \ket{\Psi(t,\theta)} = \hat{U}_\theta(t,0) \otimes \hat{\mathbb I}_\mathrm{R} \ket{\Psi(0)}.
\end{equation}
The physical state at time $t$ is recovered by tracing out the reference $ \hat{\rho}(t,\theta) = \mathrm{tr}_\mathrm{R} \{\ket{\Psi(t,\theta)} \bra{\Psi(t,\theta)}\}$ and the expectation values are obtained from
\begin{equation}
      \mathrm{tr}_\mathrm{SE} \{\hat{X}^\MRH(t,\theta) \hat{\rho}(0) \} =\bra{\Psi(t,\theta)} \hat{X} \otimes \hat{\mathbb I}_\mathrm{R} \ket{\Psi(t,\theta)}.
\end{equation}
For the purified initial state, Eq.~\eqref{eq:Alim} is an equality,
\begin{equation}
    \begin{aligned}
        \mathscr{A}^\mathrm{lim}_V(t) = \big(& \langle \partial_\theta \Psi(t,\theta)|\partial_\theta \Psi(t,\theta)\rangle\\ & -|\langle \partial_\theta \Psi(t,\theta)| \Psi(t,\theta)\rangle|^2 \big)\Big|_{\theta=1},
    \end{aligned}
\end{equation}
meaning that the limiting activity corresponds to the PDA of the purified state.
We emphasize that the limiting activity $\mathscr{A}_V^\text{lim}(t)$ does not depend on the ($\theta$-independent) purification; the reference system does not even have to be a physical system.
The PDA $\mathscr{A}_V(t)$ quantifies the sensitivity to the strength of $\hat{V}$ retained in the final physical state, while the limiting activity quantifies the sensitivity retained when comparing to a fixed reference.
Hence, $\mathscr{A}_V(t) \leq \mathscr{A}^\mathrm{lim}_V(t)$ is a consequence of the monotonicity of the QFI under CPTP maps.
For a global thermal state, for instance, the long-time rate of the PDA vanishes, while that of the limiting activity is given by the zero-frequency noise of $\hat{V}$, see App.~\ref{app:thermal}.
   
In Secs.~\ref{sec:MKUR-transport} and \ref{sec:relate} we show how the expressions~(\ref{eq:A_V},\ref{eq:Alim}) connect to previously introduced concepts of activity.

Next, to relate the response on the left hand side of~\eqref{eq:CR-KUR} to precision we consider the parametrized current operator $\hat{J}^{(X)}(t,\theta)$
\begin{equation}
\begin{split}    \label{eq:CR_response_current}
    \hat{J}^{(X)}(t,\theta) :=& \frac{i}{\hbar}[\hat{H}_\theta^\MRH(t , \theta), \hat{X}^\MRH(t,\theta)] \\
    =& \frac{i\theta}{\hbar}[ \hat{V}^\MRH(t,\theta), \hat{X}^\MRH(t,\theta)].
\end{split}
\end{equation}
Note that a change in $\theta$ changes $\hat{J}^{(X)}(t,\theta)$ in two distinct ways: It causes an explicit rescaling of the current operator (see the prefactor $\theta$ in the second line of Eq.~\eqref{eq:CR_response_current}), and it is imprinted in the entire evolution prior  to $t$ via the total system Hamiltonian.

We exploit Eq.~\eqref{eq:CR_response_current} to show how Eq.~\eqref{eq:CR-KUR} characterizes the current precision together with an additional susceptibility term $\dot{M}^{(X)}(t)$.
We write the response of the observable $\Delta \hat{X}^\MRH(t,\theta)$ to a change in the strength of $\hat{V}(t)$ as the time integral
\begin{equation}\label{eq:response-current}
    \partial_\theta \left.\langle \Delta \hat{X}^\MRH(t,\theta) \rangle\right|_{\theta=1} = 2\int_0^t ds \mathcal{J}^{(X)}(s),
\end{equation}
where the response current 
\begin{subequations}
    \begin{align}
         \mathcal{J}^{(X)}(t) :=& \frac12\left(  \langle \hat{J}^{(X)} (t) \rangle+ \Mdot\right) \label{eq: resp current}, 
\end{align}
contains not only the actual average current, $\langle \hat{J}^{(X)} (t) \rangle$, of the observable $\hat{X}$, but also the  response contribution $\Mdot$,
\begin{align}
         \Mdot :=&\frac{i}{\hbar} \!\int^t_0\!\!\!ds \langle [\hat{V}^\MRH(s), \hat{J}^{(X)} (t) ]\rangle\nonumber\\
         =& \partial_\theta\left. \left\langle \frac{\hat{J}^{(X)} (t,\theta)}{\theta} \right\rangle\right|_{\theta=1},\label{eq:M-term}
    \end{align}    
\end{subequations}
see also Ref.~\cite{Palmqvist2026} for details. While Eq.~\eqref{eq:M-term} has the shape of a susceptibility or a linear-response coefficient, we repeat that the coupling $\hat{V}(t)$ does \textit{not} have to be a small perturbation. In Eq.~\eqref{eq: resp current}, the average current originates from the response to the explicit rescaling of the current operator in Eq.~\eqref{eq:CR_response_current} while $\Mdot$ instead stems from the response of the system dynamics to the parametrized Hamiltonian.

In the limit of weak coupling, $\Mdot$ equals the current, see Appendix~\ref{app:weak-coupling-M}, and for Gorini–Kossakowski–Sudarshan–Lindblad (GKSL), dynamics, the KUR found in Ref.~\cite{Prech2025Jan} is recovered as a special case, see Appendices~\ref{app: GKSL klim} and \ref{app: SKUR GKSL}.
In general, and from a technical perspective, $\Mdot$ captures how the coupling strength $\hat{V}(s)$ at all times $s$ earlier than $t$ affects the evolution of $\hat{J}^{(X)}(t)$. Importantly, $\dot{M}^{(X)}(t)$ can be inferred by measuring the response of the current to a change in the coupling $\hat{V}(t)$, which  is often tunable, such as in fermionic~\cite{Takakura2014Mar,Wenz2019May,Thomas2020Jan,Hsiao2020May} and bosonic systems~\cite{Wulschner2016Dec,Miyanaga2021Dec,Campbell2023Jun}, see also Ref.~\cite{Fletcher2013Nov} for response measurements to a barrier-height.\footnote{In experiment it could be the case that the coupling strength is controlled by tuning a secondary parameter, i.e $\hat{H}_\theta (t) = \hat{H}_0(t) + y(\theta) \hat{V}(t)$, where we set $y(1)=1$ without loss of generality. In this case the \SKUR\ retains the same form as in~\eqref{eq:SKUR} with the only difference being a redefinition of the susceptibility term 
\begin{equation}
  \dot{M}'(t) = \frac{1}{y'(1)}\partial_\theta (J^{(X)}(t,\theta)/y(\theta) )|_{\theta=1}, \, y'(1) = \partial_\theta y(\theta)\big|_{\theta=1}, \nonumber
\end{equation}
entering the response current $\mathcal{J}^{(X)}(t) = \frac{1}{2} ( J^{(X)}(t) +\dot{M}'(t) )$, or equivalently the correction $\Phi^{(X)}(t). $}

Using the expression of Eq.~\eqref{eq:response-current} in the quantum Cram\'er-Rao bound~\eqref{eq:CR-KUR}, together with the limiting activity introduced in Eq.~\eqref{eq:Alim}, we now formulate the \SKUR\,
\begin{equation}\label{eq:SKUR}
    \frac{\left(\int_0^t ds \mathcal{J}^{(X)}(s)\right)^2}{\text{Var}[\Delta \hat{X}^\MRH(t)]} \leq  \mathscr{A}_V(t) \leq \mathscr{A}_V^\text{lim}(t),
\end{equation}
which is the central result of the paper.
In this general relation, the precision of the change of observable $\hat{X}$ generated by the coupling $\hat{V}(t)$ together with a susceptibility term is bounded by the partial dynamical activity $\mathscr{A}_V(t)$ [Eq.~\eqref{eq:A_V}], which constitutes the kinetic cost of precision.
In cases where $\mathscr{A}_V(t)$ is difficult to obtain, either theoretically or experimentally, the limiting activity $\mathscr{A}^\mathrm{lim}_V(t)$ [Eq.~\eqref{eq:Alim}] provides a more practical constraint.

Equivalently, we recast the \SKUR\  as a bound on precision 
\begin{equation}\label{eq:SKUR SNR}
    \frac{ \langle \Delta \hat{X}^\MRH(t) \rangle ^2}{\text{Var}[\Delta \hat{X}^\MRH(t)]} \leq \frac{4  \mathscr{A}_V(t)}{ \left( 1 +\Phi^{(X)}(t) \right)^2} \leq \frac{4  \mathscr{A}_V^\text{lim}(t)}{ \left( 1 +\Phi^{(X)}(t) \right)^2},
\end{equation}
where we introduced a quantum correction
\begin{equation}\label{eq:SKUR-Phi}
    \Phi^{(X)}(t) := \frac{ \int_0^t ds \dot{M}^{(X)}(s) }{\int_0^t ds \langle \hat{J}^{(X)}(s) \rangle} {=\frac{M^{(X)}(t) }{\langle \Delta \hat{X}^\MRH(t) \rangle}},
\end{equation}
with $M^{(X)}(t) := \int^t_0 ds \dot M^{(X)}(s)$. The physical current enters $ \Phi^{(X)}(t)$ in an analogous fashion as the quantum corrections found for the GKSL-limit in Refs.~\cite{Prech2025Jan,VanVu2025Mar}.

\subsection{\SKUR\ for steady-state quantum transport}\label{sec:MKUR-transport}

We now apply Eq.~\eqref{eq:SKUR} to steady-state quantum transport through out-of-equilibrium open quantum systems, which are foundational in quantum technology at large. We analyze the case where the currents reach steady-state values in the long-time limit, $\mathcal{J}^{(X)} = \lim_{t\to\infty}\mathcal{J}^{(X)}(t)$.
Taking the long-time limit in Eq.~\eqref{eq:SKUR} leads to
\begin{equation}\label{eq:SKUR-ss}
     \frac{(\mathcal{J}^{(X)})^2}{S^{(X)}} \leq  \mathscr{K}_V \leq \mathscr{K}_V^\text{lim}
\end{equation}
where
\begin{equation}\label{eq:noise}
\begin{aligned}
        S^{(X)} &= \lim_{t\rightarrow\infty} \partial_t \text{Var}[\Delta \hat{X}^\MRH(t)] \\
        &= \lim_{t\rightarrow\infty}\int_0^t ds \langle \{\delta\hat{J}^{(X)}(t),\delta\hat{J}^{(X)}(s) \}\rangle
\end{aligned}
\end{equation}
is the zero-frequency noise of $\hat{J}^{(X)}$, and $\mathscr{K}_V=\lim_{t\to\infty} \frac1t \mathscr{A}_V(t)$ is the steady-state partial dynamical activity rate. Here, for the zero-frequency noise to be well-defined, correlations need to decay in time, as it often happens, for example when a system is in contact with thermal, macroscopic reservoirs.
Likewise, the limiting activity rate $\mathscr{K}_V^\text{lim}$ is obtained in the long-time limit from Eq.~\eqref{eq:Alim},
\begin{equation}\label{eq:Klim}
\begin{split}
    \mathscr{K}_V^\text{lim} &= \lim_{t\to\infty} \frac1t \mathscr{A}^\text{lim}_V(t) = \lim_{t\to\infty} \partial_t\mathscr{A}^\text{lim}_V(t), \\
    &= \lim_{t\to\infty}\frac{1}{\hbar^2} \int_0^tds\langle \{\delta\hat{V}^\mathrm{H}(t), \delta\hat{V}^\mathrm{H}(s)\} \rangle.
\end{split}
\end{equation}
An analogous notion of an activity rate was postulated in Ref.~\cite{BlasiMay2025} in terms of the symmetrized noise in the coupling Hamiltonian. Here, we have demonstrated how the limiting activity rate $\mathscr{K}_V^\text{lim}$ stems from the quantum Fisher information, and how it limits the precision of a general quantum-transport process via the inclusion of a susceptibility term, see Eqs.~(\ref{eq:SKUR}, \ref{eq:SKUR-ss}).

Steady-state quantum transport can occur when a central system S is coupled to multiple macroscopic reservoirs, which are out of equilibrium with respect to each other. A scheme of a setup with arbitrary out-of-equilibrium reservoirs, where~(\ref{eq:SKUR}, \ref{eq:SKUR-ss}) apply is sketched in Fig.~\ref{fig: sketch general system}. 
If we are specifically interested in the precision of observables $\hat{X}_\alpha$ of reservoir $\alpha$, we replace
Eq.~\eqref{eq:Htheta} by
\begin{subequations}\label{eq: param transport ham}
\begin{align}
    \hat{H}_0 &\to \hat{H}_\text{S} + \sum_{\beta \neq\alpha} (\hat{H}_\beta + \hat{V}_\beta)+\hat{H}_\alpha \\
    \hat{V} &\to \hat{V}_\alpha, \quad\quad \hat{X} \to \hat{X}_\alpha,
\end{align}
\end{subequations}
where $\hat{H}_\text{S}$ is the Hamiltonian of the central system S, $\hat{H}_{\beta}$ is the Hamiltonian of reservoir $\beta$, and $\hat{V}_\beta$ is the coupling between S and reservoir $\beta$.
Crucially, the partial dynamical activity limiting the precision of the currents in reservoir $\alpha$ now contains the coupling $\hat{V}_\alpha^\MRH$ to reservoir $\alpha$ only, meaning that the \SKUR\ becomes a \textit{local} bound
\begin{equation}\label{eq:transport MKUR}
  \begin{aligned}
       \frac{\left(  \mathcal{J}^{(X)}_\alpha  \right)^2}{S^{(X)}_{\alpha} } 
   &\leq \mathscr{K}_\alpha \leq \mathscr{K}_\alpha^\textrm{lim}.
  \end{aligned}
\end{equation} 
Note that the assumption $[\hat{\rho}(0),\hat{X}_\alpha]=0$ only constrains the initial state with respect to $\hat{X}_\alpha$: The system and all other reservoirs $\beta\neq\alpha$ may be prepared in arbitrary, e.g., correlated or squeezed, states.

Before showing the validity and predictiveness in the concrete example of transport through a double quantum dot in Sec.~\ref{sec:dqd}, we show in Sec.~\ref{sec:relate} how the PDA and the \SKUR\ relate to previously introduced notions of dynamical activities and quantum extensions of the KUR valid in limiting cases.

\section{Connections to other notions of activity and bounds}\label{sec:relate}

Dynamical activity was first introduced in classical stochastic processes as the average number of transitions in stochastic trajectories~\cite{Garrahan2007May,Lecomte2007Apr,Maes2020Mar}.
However, in general quantum processes, a definition of dynamical activity is subtle.
For weakly coupled Markovian systems, transport can often be described in terms of quantum jumps, allowing for a natural analogue of the classical activity.
Beyond this regime, however, individual exchanges of particles and energy are not described by sharply defined jumps, since coherent scattering, strong system-reservoir hybridization, and system-environment correlations obscure the trajectory picture. This raises the question: What is a meaningful notion of dynamical activity for quantum systems?

In this section, we show that the limiting dynamical activity of Eq.~\eqref{eq:A_V} not only recovers the average number of quantum jumps in the GKSL dynamics, but also generalizes previous information-geometric notions using the quantum Fisher information~\cite{Hasegawa2020Jul,Hasegawa2023May, Nishiyama2024Apr,Hasegawa2024Dec,Yunoki2026Apr} to driven systems. Furthermore, in the weak coupling limit, we show the PDA's connection to the probability of observing changes in the outcomes of measurements~\cite{Hasegawa2021Jan,Hasegawa2025Nov,Tesser2025,Hegde2026Mar,VanVu2026May}.

\subsection{GKSL dynamics}\label{sec:relate_weak coupling}

We first show that, under the weak-coupling and Markovian approximations, the \textit{limiting} activity $\mathscr{A}^\text{lim}_V$ [Eq.~\eqref{eq:Alim}] reduces to the standard jump-counting activity, whereas the PDA $\mathscr{A}_V$ [Eq.~\eqref{eq:A_V}] is in general smaller.

Consider a system coupled to an environment,
\begin{equation}
    \hat{H}_\theta = \hat{H}_\text{S} + \hat{H}_\mathrm{E} + \theta \hat{V},
\end{equation}
where $\hat{V}$ is the coupling  Hamiltonian between system S and environment E (which possibly consists of multiple reservoirs).
Following a standard derivation of the GKSL equation~\cite{Breuer2007} and (for now) neglecting the Lamb-shift correction to the system Hamiltonian, the rescaling $\theta$ is inherited by the jump operators $\hat{L}_{k,\theta} = \theta \hat{L}_k$, and the dynamics is given by
\begin{equation}\label{eq: deformed GKSL 1}
\begin{split}
    \partial_t \hat{\rho}(t,\theta)&= \mathcal{L}_\theta \hat{\rho}(t,\theta) \\
    &= -\frac{i}{\hbar} [\hat{H}_\text{S} , \hat{\rho}(t,\theta) ] + \theta^2 \sum_k \mathcal{D}[\hat{L}_k] \hat{\rho}(t,\theta) ,
\end{split}
\end{equation}
where  $\mathcal{D}[\hat{L}_k]\hat{\rho}:= \hat{L}_k \hat{\rho} \hat{L}_k^\dagger - \frac12\{\hat{L}_k^\dagger \hat{L}_k,\hat{\rho}\}$ is the dissipator.

Evaluating Eq.~\eqref{eq:Alim} within the Born-Markov-secular approximations (see App.~\ref{app: GKSL klim}) yields the average number of jumps,
\begin{equation}
    \mathscr{A}_V^\text{lim}(t) =  \int_0^t ds \sum_k \tr{\hat{L}^\dag_k \hat{L}_k \hat{\rho}(s)},
\end{equation}
and, in the long-time limit, its rate reads
\begin{equation}
         \lim_{t\rightarrow\infty}\mathscr{K}_V^\text{lim}(t) =\lim_{t\rightarrow\infty} \partial_t \mathscr{A}^\text{lim}_V(t) =   \sum_k \tr{\hat{L}^\dag_k \hat{L}_k \hat{\rho}^\mathrm{ss}},
\end{equation}
where $\hat{\rho}^\text{ss}$ is the steady-state solution to Eq.~\eqref{eq: deformed GKSL 1}.
This is consistent with the information-geometric meaning of $\mathscr{A}^\text{lim}_V$ as the PDA of the \textit{purified} state [Eq.~\eqref{eq:purified_state}]~\cite{Gammelmark2014Apr,Prech2025Jan}: In the Markovian regime, the purified environment records every jump without statistical uncertainty.
By contrast, the \textit{partial dynamical activity} $\mathscr{A}_V$ is the QFI of the physical state $\hat{\rho}(t, \theta)$, which is in general mixed (e.g. for thermal reservoirs).
The lack of information due to the statistical uncertainty in the state enters the PDA through the factor $(\frac{p_i-p_j}{p_i+p_j})^2$, which quantifies how distinguishable the states $i$ and $j$ are in the initial ensemble.

Combining with Eq.~\eqref{eq:SKUR-ss}, the steady-state S-KUR for GKSL dynamics reads
\begin{equation}
\frac{(\mathcal J^{(X)})^2}{S^{(X)}}\leq\mathscr{K}_V\leq\mathscr{K}^{\rm lim}_V =\sum_k\text{tr}\{\hat L_k^\dagger\hat L_k\hat\rho^\text{ss}\}.
\end{equation}
When using the limiting activity rate $\mathscr{K}^{\rm lim}_V$, this is equivalent to the quantum KUR of Ref.~\cite{Prech2025Jan} (see App.~\ref{app: SKUR GKSL}); the PDA rate $\mathscr{K}_V$ instead provides a tighter bound for mixed initial states.
In this limit, the quantum correction of Eq.~\eqref{eq:SKUR-Phi} reduces to $\Phi^{(X)} = 1 + 2\psi$, with $\psi$ the correction from Ref.~\cite{Prech2025Jan}.
Since $\Phi^{(X)}$ is defined through the response to the coupling strength, this provides an operational meaning to the correction of Ref.~\cite{Prech2025Jan}.

When the Lamb shift cannot be neglected, the deformation of the coupling $\hat{V}$ also leads to a parametrization of part of the coherent dynamics [namely in the commutator of Eq.~\eqref{eq: deformed GKSL 1}].
This is expected since the Lamb shift originates from virtual system-reservoir interactions.
In turn, the quantum Fisher information no longer reduces to the average rate of jumps since it accounts for the full system-environment interaction, and not only its dissipative part, as shown in Appendix~\ref{app:Lamb-shift}.

\subsection{Total dynamical activity}\label{sec:relate_total}

The PDA of Eq.~\eqref{eq:A_V} quantifies the activity associated with a term of the Hamiltonian.
However, the generality of the decomposition Eq.~\eqref{eq:Htheta} allows us to also consider the parametrization of the full Hamiltonian.
Inspired by the classical driven case~\cite{Koyuk2020Dec,Vo2022Sep}, we consider the full Hamiltonian depending on the protocol $\lambda(t v)$ performed at speed $v$. We rescale the full Hamiltonian by substituting
\begin{equation}
\begin{split}
    \hat{H}_0(t)&\to 0,\quad \hat{V}(t)\to \hat{H}(\lambda(tv))\\
    \hat{H}_\theta(t, v) &= \theta \hat{H}(\lambda(tv))
\end{split}
\end{equation}
in Eq.~\eqref{eq:Htheta}.
The parameter $\theta$ is encoded in the state during the dynamics by the unitary
\begin{equation}\label{eq: param speed unitary}
    \begin{aligned}
        &\hat{U}_\theta(t,0;v) =T_+ \exp\left\{-\frac{i}{\hbar}\int_0^{t}\!\! ds \theta \hat{H}(\lambda(s v)) \right\}\\
        &= T_+\exp\left\{-\frac{i}{\hbar}\int_0^{t\theta} \!\!ds \hat{H}(\lambda(s \frac{v}{\theta})) \right\} = \hat{U}(\theta t,0 ;\frac{v}{\theta}).
    \end{aligned}
\end{equation}
From this parametrization, the total dynamical activity $\mathscr{A}_H(t,v)$ and its limiting activity $\mathscr{A}^\text{lim}_H(t,v)$ are obtained by replacing $\hat{V}(t)\to \hat{H}(t,v)$ in Eqs.~\eqref{eq:A_V} and \eqref{eq:Alim}, respectively.

The introduction of the protocol speed $v$ allows us to write the response of an operator to changing $\theta$ as
\begin{equation}
\begin{split}
    \partial_\theta &\tr{\hat{X}^\mathrm{H}(\theta t,\frac{v}{\theta}) \hat{\rho}(0) } \Big|_{\theta =1}=\\
    &\qquad\qquad=(t\partial_t - v\partial_v) \tr{\hat{X}^\mathrm{H}( t,v ) \hat{\rho}(0) }.
\end{split}
\end{equation}
Then, the quantum Cram\'er-Rao bound~\eqref{eq:CR-KUR} becomes
\begin{equation}\label{eq: quantum KUR protocol}
    \frac{ \left((t\partial_t - v\partial_v) \langle\hat{X}^\mathrm{H}( t,v)  \rangle \right)^2}{\text{Var} [\hat{X}^\mathrm{H}(t,v)]} \leq 4 \mathscr{A}_H(t,v)\leq 4 \mathscr{A}_H^\mathrm{lim}(t,v),
\end{equation}
which is the quantum analogue of the driven classical KUR, where the differential operator $(t\partial_t-v\partial_v)$ appears for time-dependent transition rates~\cite{Koyuk2020Dec,Vo2022Sep}. 
For the time-independent case, the $v\partial_v$ contribution vanishes as $\lambda(tv)=\lambda$, and integrating the limiting activity Eq.~\eqref{eq:Alim} leads to
\begin{equation}\label{eq: time indep QKUR}
        \frac{ \left(t\partial_t\langle\hat{X}^\mathrm{H}( t)  \rangle \right)^2}{\text{Var} [\hat{X}^\mathrm{H}(t)]} \leq 4 \mathscr{A}_H(t)\leq 4\frac{t^2}{\hbar^2}\text{Var}[\hat{H}].
\end{equation}
This inequality is equivalent to the observable-level form of the \emph{time-energy uncertainty relation} underlying the Mandelstam-Tamm quantum speed limit~\cite{Mandelstam1945, Deffner2017Oct}, as also noted in Refs.~\cite{Hasegawa2023May,Nakajima2023Nov, Nishiyama2024Apr, Nishiyama2025Apr}.
Here, the $\theta$-parametrization is equivalent to a parametrization of time $t$, whereas in the driven case they lead to different quantum Fisher information, as shown in Appendix~\ref{app:total-DA}.

The time-independent case also showcases the difference between the PDA and the limiting activity: Consider an initial state diagonal in the energy eigenbasis $[\hat{\rho}(0),\hat{H}]=0$. In this case, the PDA vanishes $\mathscr{A}_H(t)=0$, reflecting the fact that the system undergoes no evolution $\hat{\rho}(t)=\hat{\rho}(0)$.
Instead, $\mathscr{A}^\mathrm{lim}_H(t) \neq 0$ in general since it is determined by the initial statistical uncertainty of $\hat{\rho}(0)$.
Therefore, while the limiting activity is more straightforwardly connected to the average number of jumps for open quantum systems [see Appendix~\ref{app: GKSL klim}], it does not quantify the speed of evolution of the physical state, which is instead captured by the PDA.

However, Eq.~\eqref{eq: time indep QKUR} is challenging to apply to analyze current precision in a transport setting because (i) $\text{Var}[\hat{H}]$ is very large due to the size of the macroscopic reservoir, and (ii) in the long-time limit Eq.~\eqref{eq: time indep QKUR} is trivial because of the quadratic scaling in time of $\mathscr{A}_H(t)$. 
By contrast, the partial and its limiting dynamical activities relevant for transport [Eqs.~(\ref{eq:A_V}, \ref{eq:Alim})] are expressed in terms of correlation functions of the coupling Hamiltonian $\hat{V}$. In the setting where a quantum system is coupled to large reservoirs, these correlation functions typically decay, leading to linear scaling of the partial dynamical activities for long times.

\subsection{Measurement-based activity}\label{sec:relate_measurement based}

Another operational route to quantum dynamical activity is based on the classical probability distribution generated by performing measurements on a quantum system.
Suppose that a measurement protocol produces a record $\varphi= \{\varphi_1,\dots,\varphi_N \}$ with probability $P(\varphi)$.
For each outcome in the measurement record, the outcomes of the observables are encoded in the stochastic variable $\breve{X}(\varphi_i)$.
The subset of trajectories where no change in the observable outcome is detected, $\Upsilon = \{\varphi| \breve{X}(\varphi_{1})=\breve{X}(\varphi_{q})\,\forall q\in{1,\dots, N}\}$, has probability measure
\begin{equation}\label{eq:survival-P}
    P_0 = \sum_{\varphi \in \Upsilon} P(\varphi) \leq 1,
\end{equation}
and has been shown to enter general bounds on fluctuations in Refs.~\cite{Hasegawa2021Jan,Hasegawa2025Nov,Tesser2025,Hegde2026Mar,VanVu2026May}.
In particular, $1-P_0$ quantifies the probability of seeing a change in $\breve{X}(\varphi_i)$, and has been referred to as survival-activity~\cite{Hasegawa2021Jan}.

The partial dynamical activity Eq.~\eqref{eq:A_V} and its limiting activity Eq.~\eqref{eq:Alim} are also connected to the measurement-based approach.
To see this, we introduce the conditional probability $P(i|j) = |\braket{i|\hat{U}(t,0)|j}|^2$ of observing a transition from state $\ket{j}$ to $\ket{i}$ after the system has evolved for time $t$.
We focus on time-reversal symmetric systems, for which $P(i|j)= P(j|i)$, as is commonly assumed in uncertainty relations derived from fluctuation theorems~\cite{Hasegawa2019Sep, Timpanaro2019Aug}, and write Eq.~\eqref{eq:A_V} as
\begin{equation}\label{eq:A_V-sigma}
\begin{split}
    \mathscr{A}_V(t) &=\sum_{ij} \tanh^2\left(\frac{\sigma_{ij}}{2}\right) \frac{|\braket{i|\int_0^t\!\!\frac{ds}{\hbar}\hat{V}^\MRH(s)|j}|^2}{P(i|j)}\times\\
    &\qquad\times\frac{P(j;i)+P(i;j)}{2},
\end{split}
\end{equation}
where we introduced the stochastic entropy production
\begin{equation}
    \sigma_{ij}:= \log \frac{P(i;j)}{P(j;i)},\, P(i;j)= p_j P(i|j).
\end{equation}
Time-reversal symmetry is essential here to link the trajectory formulation of the stochastic entropy production with the statistics of the initial ensemble.
In the weak-coupling limit, $t\hat{V}/\hbar\ll1$, the conditional probability $P(i|j)\approx |\braket{i|\int_0^t\!\!\frac{ds}{\hbar}\hat{V}^\MRH(s)|j}|^2$ for $i\neq j$, which are the only contributions to the sum of Eq.~\eqref{eq:A_V-sigma}.
In this limit, the PDA therefore reads
\begin{equation}
\begin{split}
    \mathscr{A}_V(t)&\approx \sum_{ij} \left[\tanh\left(\frac{\sigma_{ij}}{2}\right)\right]^2 \frac{P(i;j)+P(j;i)}{2},\\
    &= \sum_{ij} \tanh\left(\frac{\sigma_{ij}}{2}\right) P(i;j),
\end{split}
\end{equation}
and is analogous to the \textit{thermodynamic} constraint on precision established from fluctuation theorems~\cite{Hasegawa2019Sep, Timpanaro2019Aug, Potts2019Nov}.
The PDA is bounded using $\tanh^2(\sigma_{ij}/2)\leq 1$ for $i\neq j$ and $\sigma_{ii}=0$,
\begin{equation}
\begin{split}
  \mathscr{A}_V(t)&\leq \sum_{i, j\neq i}P(i; j) = 1-\sum_i P(i;i)\leq \mathscr{A}_V^\mathrm{lim}(t),
\end{split}
\end{equation}
where the last inequality becomes an equality when $\hat V$ has no diagonal elements in the eigenbasis of $\hat\rho(0)$.
This expression has the same form as the survival activity $1-P_0$. Importantly, it shows that the \SKUR\ developed in this paper not only includes kinetic but also thermodynamic aspects of the dynamics.

\section{\SKUR\ applied to quantum transport}\label{sec:dqd}
To illustrate the \SKUR\ in quantum transport, we use a double quantum dot (DQD) connected to two macroscopic reservoirs.
\subsection{Double quantum dot coupled to macroscopic reservoirs: Model and transport observables}\label{subsec:modelDQD_baths}
The Hamiltonian of the double-dot system, assuming spinless electrons, is
\begin{equation}\label{eq:H_DQD}
    \hat{H}_\mathrm{S} = \epsilon_1 \hat{d}^\dag_1 \hat{d}_1 +\epsilon_2 \hat{d}^\dag_2 \hat{d}_2 +g (\hat{d}^\dag_1 \hat{d}_2 +\hat{d}^\dag_2 \hat{d}_1),
\end{equation}
where $\epsilon_i$ is the single-particle on-site energy and $\hat{d}^\dag_i$ the creation operator of the electrons in dot $i$. The coherent tunneling strength between the dots is given by $g$. From now on, we set $\epsilon_1=\epsilon_2 =\epsilon_d$  and the eigenenergies of $\hat{H}_\MRS$ are hence $\epsilon_{\pm} = \epsilon_d \pm g$. The Hamiltonian of reservoir $\alpha$ is
\begin{equation}
    \hat{H}_\alpha =\sum_{k} \epsilon_{\alpha k} \hat{c}_{\alpha k}^\dag \hat{c}_{\alpha k},
\end{equation}
and the coupling between the system and wide-band reservoirs are given by the tunneling Hamiltonians
\begin{equation}
\begin{aligned}
        \hat{V}_\MRL &= \sum_k  (t_{ \MRL k 1} \hat{c}^\dag_{\MRL k} \hat{d}_1 + t^*_{\MRL k 1} \hat{d}^\dag_1 \hat{c}_{\MRL k}),\\
        \hat{V}_\MRR &= \sum_k  (t_{ \MRR k 2} \hat{c}^\dag_{\MRR k} \hat{d}_2 + t^*_{\MRR k 2} \hat{d}^\dag_2 \hat{c}_{\MRR k} ).
\end{aligned}
\end{equation}
Here, $\hat{c}^\dag_{\alpha k}$ is the creation operator of electrons in the $k$-th energy state of reservoir $\alpha$. 

In the following, we demonstrate the high predictiveness of the limiting \SKUR,
\begin{equation}\label{eq: ss transport relax MKUR}
   {\mathscr{P}}_{\alpha}^{(X)} :=\frac{\left(  \mathcal{J}^{(X)}_\alpha  \right)^2}{S^{(X)}_{\alpha} } =\frac{\left(  {J}^{(X)}_\alpha +\dot{M}_\alpha^{(X)}  \right)^2}{4S^{(X)}_{\alpha} }
  \leq \mathscr{K}_\alpha^\mathrm{lim},
\end{equation}
for coherent charge and energy transport through a double dot and compare it to the local KUR valid for classical, noninteracting particles~\cite{Palmqvist2025Oct,Palmqvist2026}\footnote{The general classical KUR would use the activity of both reservoirs $(J^{(X)}_\alpha)^2 /S^{(X)}_\alpha \leq \sum_\beta \mathscr{K}^\mathrm{lim}_\beta=:\mathscr{K}^\mathrm{lim}$, but is still broken in the DQD~\cite{Prech2025Jan}.}
\begin{equation}\label{eq:classical-KUR}
{P}_{\alpha}^{(X)} :=\frac{(J^{(X)}_\alpha)^2}{S^{(X)}_\alpha} \leq \mathscr{K}^\mathrm{lim}_\alpha\ .
\end{equation}

We now provide the explicit expressions for all quantities entering the \SKUR~\eqref{eq: ss transport relax MKUR}. 
As discussed in Sec.~\ref{sec:MKUR-transport}, we parametrize the full Hamiltonian as 
\begin{equation}
    \hat{H}_\theta = \hat{H}_\text{S} + \hat{H}_\MRL +\hat{H}_\MRR + \theta \hat{V}_\MRL +\hat{V}_\MRR,   
\end{equation}
defining a partial dynamical activity with respect to the coupling with the left reservoir $\mathscr{A}_\MRL(t) := \mathscr{A}_{V_\MRL}(t)$.
The steady-state particle current depends on $\theta$ through the parametrization imprinted on the state evolution, 
\begin{equation}
    J^{(N)}_\MRL(\theta) = \lim_{t\rightarrow\infty} \partial_t \langle \hat{N}^\MRH_\MRL(t,\theta) \rangle,\, \hat{N}_\MRL=\sum_k \hat{c}_{\MRL k}^\dag \hat{c}_{\MRL k}.
\end{equation}
Since the Hamiltonian in Eq.~\eqref{eq:H_DQD} is quadratic, the analytical expression of the parametrized average current follows as~\cite{Meir1992Apr}
\begin{equation}\label{eq:particlecurrent}
    J^{(N)}_\MRL(\theta) =\frac{1}{\hbar} \int^\infty_{-\infty}  \frac{d \epsilon}{2 \pi} \mathcal{T}_{\MRL\MRR}(\epsilon,\theta) (f_\MRR (\epsilon) -f_\MRL(\epsilon)),
\end{equation}
where $f_\alpha (\epsilon) := 1/(e^{(\epsilon -\mu_\alpha)/\kB T_\alpha}+1)$ is the Fermi function of reservoir $\alpha$ with temperature $T_\alpha$ and chemical potential $\mu_\alpha$. Analogous expressions are found for the energy current, by replacing $\hat{N}_\MRL\to\hat{H}_\mathrm{L}$ and consequently introducing a factor $\epsilon$ in the integrand of Eq.~\eqref{eq:particlecurrent}, see also Appendix~\ref{app: SKUR energy DQD}.
The parameter $\theta$ appears in the transmission function $\mathcal{T}_{\MRL\MRR}(\epsilon,\theta)$, which is obtained by means of nonequilibrium Green's functions and reads~\cite{Caroli1971Jun,Covito2018May,BlasiMay2025}
\begin{equation}
   \begin{aligned}
        &\mathcal{T}_{\MRL\MRR}(\epsilon ,\theta) = \\
    &= \frac{ g^2 \theta^2\Gamma_\mathrm{L} \Gamma_\mathrm{R}}{ (g^2 + \frac{\theta^2\Gamma_\mathrm{L} \Gamma_\mathrm{R} }{4} -(\epsilon-\epsilon_d)^2)^2 +\frac{(\theta^2\Gamma_\mathrm{L}+ \Gamma_\mathrm{R})^2}{4} (\epsilon-\epsilon_d)^2} ,
   \end{aligned}
\end{equation}
where the spectral functions $\Gamma_\alpha$ are energy-independent due to the wide-band assumption.
The effect of rescaling $\hat{V}_\MRL \mapsto \theta \hat{V}_\MRL$ is $\Gamma_\MRL \mapsto \theta^2\Gamma_\MRL$. For conciseness, when $\theta=1$ we drop the $\theta$-argument. From the $\theta$-dependent particle or energy currents, the susceptibility $\dot{M}_\MRL^{(X)}$ is obtained according to Eq.~\eqref{eq:M-term}. 

The zero-frequency noise in the particle current is~\cite{blanter_shot_2000}
\begin{equation}
\begin{aligned}
        S^{(N)}_\mathrm{L}&= \lim_{t\rightarrow\infty} \partial_t \mathrm{Var}[\hat{N}^\mathrm{H}_{\color{red}\MRL}(t)-\hat{N}^\mathrm{H}_{\color{red}\MRL}(0)]\\
    &=\frac{1}{\hbar} \int^{\infty}_{-\infty}\frac{d\epsilon}{2\pi}\left\{  \mathcal{T}_{\MRL\MRR}(\epsilon) (F_{\MRL\MRR}(\epsilon) +F_{\MRR\MRL}(\epsilon) )\right\} \\
    &\quad-\frac{1}{\hbar} \int^{\infty}_{-\infty}\frac{d\epsilon}{2\pi}\left\{  \mathcal{T}_{\MRL\MRR}(\epsilon) (f_{\MRR}(\epsilon) -f_{\MRL}(\epsilon) )\right\}^2,
\end{aligned}
\end{equation}
where we introduced $F_{\alpha \beta}(\epsilon) := f_\alpha(\epsilon) (1-f_\beta(\epsilon))$. Again, the energy current noise has an additional factor $\epsilon^2$ in the integrands.

As mentioned in Sec.~\ref{sec:MKUR}, while progress can be made in the linear-response regime~\cite{Mihailescu2024Jun}, a direct calculation of the partial dynamical activity $\mathscr{A}_\text{L}(t)$ [Eq.~\eqref{eq:A_V}] is not tractable in a general nonequilibrium setting, whereas the limiting activity $\mathscr{A}_\text{L}^\text{lim}(t)$ [Eq.~\eqref{eq:Alim}] is, both for fermionic transport, as shown in the following, and for bosonic transport through nonlinear systems~\cite{Palmqvist2026Aug}.
In particular, the limiting activity rate $\mathscr{K}^\text{lim}_\text{L}$ [Eq.~\eqref{eq:Klim}] reads~\cite{BlasiMay2025}
\begin{equation}\label{eq:Klim_scattering}
    \begin{aligned}
        &\mathscr{K}^\mathrm{lim}_\mathrm{L} = \frac{1}{\hbar} \int^{\infty}_{-\infty}\frac{d\epsilon}{2\pi}\left\{  \mathcal{T}_{\MRL\MRR}(\epsilon) (F_{\MRL\MRR}(\epsilon) +F_{\MRR\MRL}(\epsilon) )\right\} +\\
        &+\frac{1}{\hbar} \int^{\infty}_{-\infty}\frac{d\epsilon}{2\pi} [4\mathcal{T}_{\MRL\MRL}(\epsilon)-(\mathcal{T}_{\MRL\MRL}(\epsilon) +\mathcal{T}_{\MRL\MRR}(\epsilon))^2] F_{\MRL\MRL}(\epsilon).
    \end{aligned}
\end{equation}
 Here, in addition to $\mathcal{T}_{\MRL\MRR}(\epsilon)$ also
\begin{equation}
        \mathcal{T}_{\MRL\MRL}(\epsilon)=\frac{\Gamma_\MRL^2 \left(\frac{\Gamma_\MRR^2}{4}+(\epsilon-\epsilon_d)^2\right)}{\left(\frac{\Gamma_\MRL \Gamma_\MRR}{4}+g^2-(\epsilon-\epsilon_d)^2\right)^2+\frac{(\Gamma_\MRL+\Gamma_\MRR)^2}{4}  (\epsilon-\epsilon_d)^2}
\end{equation}
enters~\cite{BlasiMay2025} due to thermal fluctuations in the left reservoir. In the limit of $T_\mathrm{L, R}\to0$, the second line of Eq.~\eqref{eq:Klim_scattering} vanishes and we find
\begin{equation}
\begin{aligned}
    \mathscr{K}^\mathrm{lim}_\mathrm{L} &= \frac{1}{\hbar} \int^{\infty}_{-\infty}\frac{d\epsilon}{2\pi}\left\{  \mathcal{T}_{\MRL\MRR}(\epsilon) (F_{\MRL\MRR}(\epsilon) +F_{\MRR\MRL}(\epsilon) )\right\}\\
       &=\frac{1}{\hbar} \int^{\Delta \mu/2}_{-\Delta\mu/2}\frac{d\epsilon}{2\pi} \mathcal{T}_{\MRL\MRR}(\epsilon) = \left|J^{(N)}_\MRL \right|
\end{aligned}
\end{equation}
In the zero-temperature limit, the activity is hence directly measurable via a current measurement. This is practically relevant as soon as thermal fluctuations in the left reservoir can be neglected.

\begin{figure}
    \centering
   \includegraphics{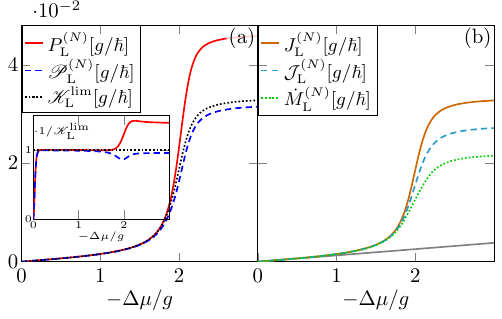}
\caption{Classical local KUR [Eq.~\eqref{eq:classical-KUR}] and \SKUR\ [Eq.~\eqref{eq: ss transport relax MKUR}] for particle current in DQD as a function of chemical potential bias $\Delta \mu$. We set $\Gamma_\MRR = 5\Gamma_\MRL =g/5 $, $\ T_\MRL= T_\MRR =  g/50\kB$ and $\epsilon_d=0$. (a) Violation of local classical KUR and tightness of the \SKUR. (b) Particle current $J^{(N)}_\MRL$, susceptibility $\dot{M}_\text{L}^{(N)}$ and their average $\mathcal{J}^{(N)}_\MRL$. The close-to-equilibrium approximation $\dot{M}_\text{L}^{(N)}\approx -\frac{\Gamma_\MRL\Gamma_\MRR}{h g^2} \Delta\mu$ of Eq.~\eqref{eq:M_limit} (valid for $g\gg \Delta\mu,\Gamma_{\MRL,\MRR}$) is shown in gray. }
    \label{fig: DQD intro}
\end{figure}

\subsection{Validity and predictiveness of S-KUR and classical KURs}

We show the results for the \SKUR~\eqref{eq: ss transport relax MKUR} and the noninteracting classical limit Eq.~\eqref{eq:classical-KUR} in Fig.~\ref{fig: DQD intro} for the particle current.
The classical local KUR [Eq.~\eqref{eq:classical-KUR}] is violated when the bias $|\Delta\mu|=|\mu_\MRL - \mu_\MRR|$ reaches the level splitting $2g$, see Fig.~\ref{fig: DQD intro}(a), and both eigenstates of the DQD lie in the transport window around $\bar{\mu}=\left(\mu_\MRL+\mu_\MRR\right)/2$.
While previous violations of the classical, noninteracting KUR~\eqref{eq:classical-KUR} were shown in classical dynamics in the presence of Coulomb interactions~\cite{Monsel2026May} and in superconducting hybrid devices~\cite{Manzano2023Oct}, we here treat noninteracting electrons, and the violation of Eq.~\eqref{eq:classical-KUR} is fully due to quantum effects. 
Instead, the \SKUR~\eqref{eq: ss transport relax MKUR} is a tight bound in a broad range of $\Delta\mu$, with a dip in tightness occurring when the classical bound's violation becomes substantial.
Up until this point in $\Delta\mu$, $J^{(N)}_\MRL \approx \dot{M}_\MRL^{(N)}$ as shown in Fig.~\ref{fig: DQD intro}(b), meaning that the \SKUR\ is equivalent to the classical KUR.
This happens because, when the DQD levels lie far from the transport window ($2g\gg\Delta\mu$) and $g^2\gg \Gamma_\MRL\Gamma_\MRR$ (as in Fig.~\ref{fig: DQD intro}), the transmission function in the transport window approximates as $\mathcal{T}_\text{LR}(\epsilon,\theta)\approx \theta^2 \Gamma_\MRL\Gamma_\MRR/g^2$, and both current and susceptibility read 
\begin{equation}\label{eq:M_limit}
J^{(N)}_\MRL \approx \dot{M}_\MRL^{(N)}\approx - \frac{\Gamma_\MRL\Gamma_\MRR}{h g^2} \Delta\mu,
\end{equation}
shown in gray in Fig.~\ref{fig: DQD intro}(b).
The same $\theta^2$-dependence of the transmission function is also obtained in the weak-coupling limit $\Gamma_\MRL\to 0$, and results in $J^{(N)}_\MRL \approx \dot{M}_\MRL^{(N)}$, see Appendix~\ref{app:weak-coupling-M} and~\ref{app: SKUR GKSL}.
At large $\Delta\mu$, where the classical KUR is violated, the \SKUR\ remains tight instead.
Consequently, 
\begin{equation}
    \frac{\left(\mathcal{J}^{(N)}_\mathrm{L} \right)^2}{ S^{(N)}_\mathrm{L}} \approx \mathscr{K}_\MRL \approx \mathscr{K}^\mathrm{lim}_\MRL. 
\end{equation}
This also has an information-geometric implication: Since the particle current saturates the \SKUR, it means that, for this regime, $\hat{N}_\MRL$ is close to an optimal observable for estimating the coupling strength between the double dot and the left reservoir.

\begin{figure}
    \centering
   \includegraphics{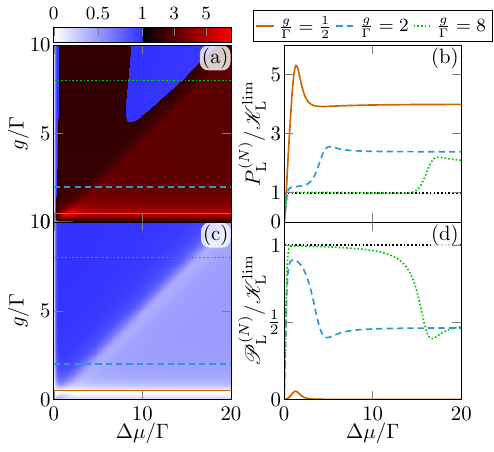}
\caption{(a,b) Violation of classical local KUR [Eq.~\eqref{eq:classical-KUR}] and (c,d) tightness of \SKUR\ [Eq.~\eqref{eq: ss transport relax MKUR}] in DQD for varying tunneling strength $g$ and chemical potential bias $\Delta \mu$. We set $\Gamma_\MRL=\Gamma_\MRR =\Gamma$, $T_\MRL =T_\MRR =0.1 \Gamma/\kB$, and $\epsilon_d=0$.}
    \label{fig: DQD g mu}
\end{figure}

We now analyze in more detail the characteristics of precision and its bounds for both charge and energy currents and for a broad range of parameter regimes. In Fig.~\ref{fig: DQD g mu}, we illustrate the violation of the classical local KUR and the tightness of the \SKUR\ when varying the coherent coupling strength $g$ and the chemical potential bias $\Delta \mu$ for symmetric tunnel couplings $\Gamma_\MRL = \Gamma_\MRR = \Gamma$.
Substantial violation of the classical bound is shown in Fig.~\ref{fig: DQD g mu}(a,b), reaching a  maximum for $2g\approx \Gamma$, where the hopping time between the dots matches the dwell time in the dots.
For this value of $g$ the \SKUR\ is far from being saturated, see Fig.~\ref{fig: DQD g mu}(c,d), where $J^{(N)}_\alpha \approx-\dot{M}_\alpha^{(N)} \Rightarrow \mathcal{J}^{(N)}_\MRL\approx 0$. This means that the accumulated number of particles in the reservoir does not respond to a change in the coupling strength in this regime.
Instead, the \SKUR\ becomes tight at large values of $g$ when the levels lie outside of the bias window, $2g>\Delta\mu$. See Appendix~\ref{app: SKUR energy DQD} for additional plots.

\begin{figure}
    \centering
   \includegraphics{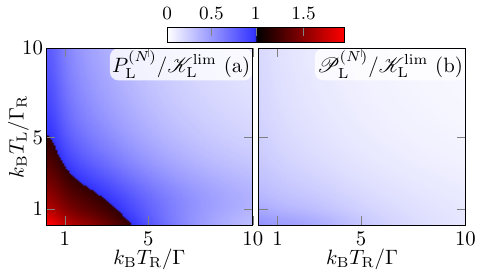}
\caption{
(a) Violation of classical local KUR [Eq.~\eqref{eq:classical-KUR}] and (b) tightness of \SKUR\ [Eq.~\eqref{eq: ss transport relax MKUR}] in DQD for varying temperatures $T_\MRL,T_\MRR$. We set $\Gamma_\MRL=\Gamma_\MRR =\Gamma$, $g=5\Gamma$, $\epsilon_d=2\Gamma$, $\Delta \mu=10 \Gamma$. }
    \label{fig: DQD temperature}
\end{figure}

In Figs.~\ref{fig: DQD intro} and \ref{fig: DQD g mu}, violation of the classical local KUR and tightness of the \SKUR\ are shown at low temperatures. 
When the temperatures are increased, $k_\text{B}T_\mathrm{L/R}>\Gamma$, as shown in Fig.~\ref{fig: DQD temperature}, the violations of the classical bound disappear [Fig.~\ref{fig: DQD temperature}(a)], and the \SKUR\ is not saturated any longer. 
Indeed, by increasing the temperature, not only is the particle current noise increased (reducing the precisions $P_\MRL^{(N)},\mathscr{P}_\MRL^{(N)}$), but also the limiting activity $\mathscr{K}^\mathrm{lim}_\mathrm{L}$ increases as it contains more thermal fluctuations in addition to the fluctuations related to parameter sensitivity.

\begin{figure*}
    \centering
    \includegraphics[width=6.6in]{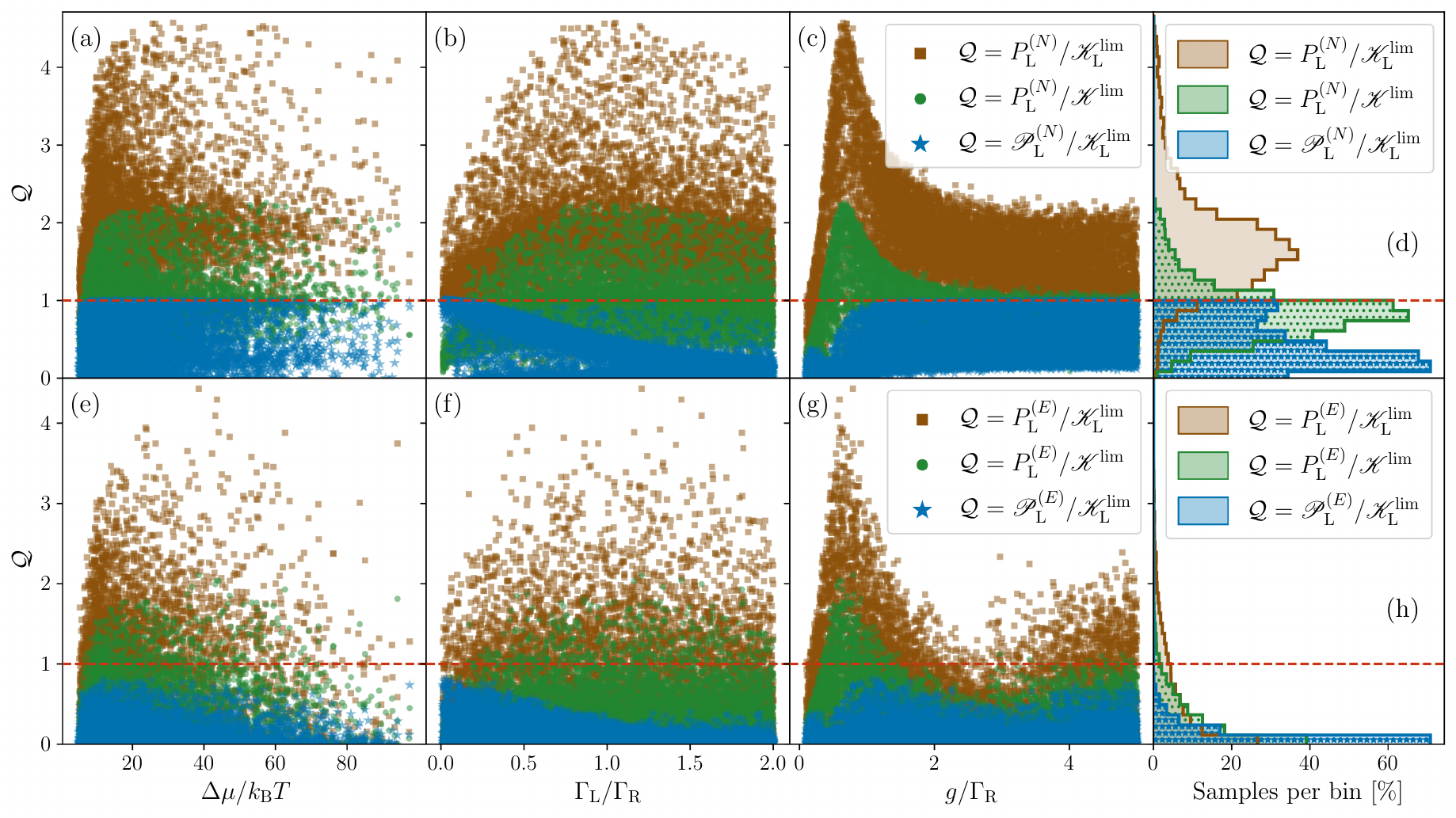}
    \caption{Ratio between relevant precisions and limiting activities for the S-KUR, as well as local and global classical KURs. Plots demonstrate how tight the \SKUR\ is and when violations of the classical KURs occur for (a,b,c,d) the particle current and (e,f,g,h) the energy current. We have sampled $10000$ points uniformly in the range $\Gamma_\MRL \in [\Gamma_\MRR/100,2\Gamma_\MRR]$, $\kB T_\MRL=\kB T_\MRR\in [\Gamma_\MRR/100, \Gamma_\MRR ]$, $\Delta\mu\in[5 \Gamma_\MRR,10\Gamma_\MRR]$, $\bar{\mu}\in[-2\Gamma_\MRR,2\Gamma_\MRR]$, $g\in[\Gamma_\MRR/10,5\Gamma_\MRR]$, $\epsilon_d \in[-3\Gamma_\MRR ,3\Gamma_\MRR]$.
    }  
    \label{fig:SKUR sample}
\end{figure*}

Finally, to get an overview of how tight the \SKUR\ in the DQD  is for both charge and energy currents in different parameter regimes and when local and total classical KURs are broken, we sample over a large range of parameter points in Fig.~\ref{fig:SKUR sample}, where we display the factor given by the ratio between the relevant precision and the limiting activity for the particle and energy currents. 

The particle and energy-current precisions display rather different behavior:  While the local and the classical KURs are violated for both current precisions, these violations are more prominent for the particle current. Moreover, the \SKUR\ is in general tighter for the particle-current precision than for the energy-current precision. We note that for large coupling asymmetry, $\Gamma_\mathrm{L}/\Gamma_\mathrm{R}\ll1$, the \SKUR\ is closest to saturation---both in particle- and energy-current precisions.  

That the coupling asymmetry $\Gamma_\MRL/\Gamma_\MRR$ plays a key role in how tight the \SKUR\ is is to be expected, since this asymmetry is closely related to how sensitive a quantity is to a variation of the coupling strength $\Gamma_\MRL$. If the coupling strength to the left reservoir is small, it acts as a bottleneck limiting the transport. Small changes in $\Gamma_\MRL$ then change the transport currents significantly if the system is out of equilibrium. This translates into large sensitivities to changes in $\Gamma_\MRL$ and hence a tight \SKUR.

The DQD results of this section demonstrate that the \SKUR\ provides a tight bound on precision at steady-state, a regime which is highly relevant for quantum transport, quantum thermodynamics, and quantum technologies.
While the classical KUR bound based on just the current, Eq.~\eqref{eq:classical-KUR}, can fail in coherent quantum transport, the response current $\curlyJ$ in Eq.~\eqref{eq: ss transport relax MKUR} obeys the universal bound.
Moreover, the \SKUR\ is not limited to just the particle current, but holds for general transport observables.

\subsection{Comparison with bounds on particle-current precision in noninteracting systems}
In this section we compare the S-KUR with recently developed bounds on the particle-current precision in coherent quantum transport in noninteracting systems. To make the comparison clearer we cast the S-KUR as an upper bound on the current precision, taken the special case of  Eq.~\eqref{eq:SKUR SNR} for $X\to N_\MRL$,
\begin{equation}\label{eq: SKUR P bound SS}
    \frac{(J^{(N)}_\MRL)^2}{S^{(N)}_\MRL} \leq \frac{4 \mathscr{K}^\mathrm{lim}_\MRL}{ \left( 1 +\Phi_\MRL^{(N)} \right)^2} := \mathcal{B}_\mathrm{SKUR} .
\end{equation}
We emphasize that the appearance of the current in the upper bound of~\eqref{eq: SKUR P bound SS} is similar to how quantum corrections to the KUR and TKUR were formulated in  Refs.~\cite{Prech2025Jan,VanVu2025Mar} for GKSL dynamics.

We compare the S-KUR with the quantum TUR for particle transport discovered in Ref.~\cite{Brandner2025Jul},
\begin{equation}\label{eq: QTUR}
     \frac{(J^{(N)}_\MRL)^2}{S^{(N)}_\MRL} \leq J^{(N)}_\MRL \sinh\left[\frac{\sigma}{2 \kB J^{(N)}_\MRL } \right] := \mathcal{B}_\sigma,
\end{equation}
where $\sigma$ is the average entropy production rate.
In addition, we compare the S-KUR with the recently proposed precision bound of Ref.~\cite{BlasiMay2025} 
\begin{equation}\label{eq: QKUR}
    \frac{(J^{(N)}_\MRL)^2}{S^{(N)}_\MRL} \leq \frac{(\mathscr{K}^\mathrm{cross}_\MRL)^2}{\mathscr{K}^\mathrm{cross}_\MRL -\mathscr{K}^\mathrm{sh}_\MRL} =: \mathcal{B}_\mathrm{cross}.
\end{equation}
Here, since we assume time-translation invariance, the correlator-based definition of activity of Ref.~\cite{BlasiMay2025} coincides with the limiting activity defined in Eq.~\eqref{eq:Klim}. The quantities entering the upper bound of~\eqref{eq: QKUR} are 
\begin{equation}
\begin{aligned}
        \mathscr{K}^\mathrm{cross}_\MRL &=\frac{1}{\hbar} \int^{\infty}_{-\infty}\frac{d\epsilon}{2\pi}\left\{  \mathcal{T}_{\MRL\MRR}(\epsilon) (F_{\MRL\MRR}(\epsilon) +F_{\MRR\MRL}(\epsilon) )\right\}\\
          \mathscr{K}^\mathrm{sh}_\MRL &=\frac{1}{\hbar} \int^{\infty}_{-\infty}\frac{d\epsilon}{2\pi}\left\{  \mathcal{T}_{\MRL\MRR}(\epsilon) (f_{\MRR}(\epsilon) -f_{\MRL}(\epsilon) ) ^2\right\}\ .
\end{aligned}
\end{equation}
They originate from two different mathematical decompositions of the limiting activity 
\begin{equation}
    \mathscr{K}^\mathrm{lim}_\MRL = \mathscr{K}^\mathrm{auto}_\MRL+\mathscr{K}^\mathrm{cross}_\MRL = \mathscr{K}^\mathrm{th}_\MRL+\mathscr{K}^\mathrm{sh}_\MRL,
\end{equation}
for detailed expressions of $\mathscr{K}^\mathrm{th}_\MRL$ and $\mathscr{K}^\mathrm{auto}_\MRL$ see Ref.~\cite{BlasiMay2025}. While it is possible to make interpretations of the processes contained in each of the different terms used to split $\mathscr{K}^\mathrm{lim}_\MRL$, we emphasize that such a decomposition is not unique and the individual terms on their own lack the rigorous microscopic and information geometric definition which $\mathscr{K}^\mathrm{lim}_\MRL$ possesses. Furthermore, we note that the denominator entering the cost function of~\eqref{eq: QKUR} can be written as $\mathscr{K}^\mathrm{cross}_\alpha -\mathscr{K}^\mathrm{sh}_\alpha = S^{(N)\mathrm{th}}_\alpha$, with the thermal noise~\cite{blanter_shot_2000}
\begin{equation}
    \begin{aligned}
        S^{(N)\mathrm{th}}_\MRL &=\int^{\infty}_{-\infty}\frac{d\epsilon}{2\pi\hbar} \mathcal{T}_{\MRL\MRR}(\epsilon) (F_{\MRL\MRL}(\epsilon) +F_{\MRR\MRR}(\epsilon) ).
 \end{aligned}
\end{equation}

\begin{figure}
    \centering
   \includegraphics[width=3.3in]{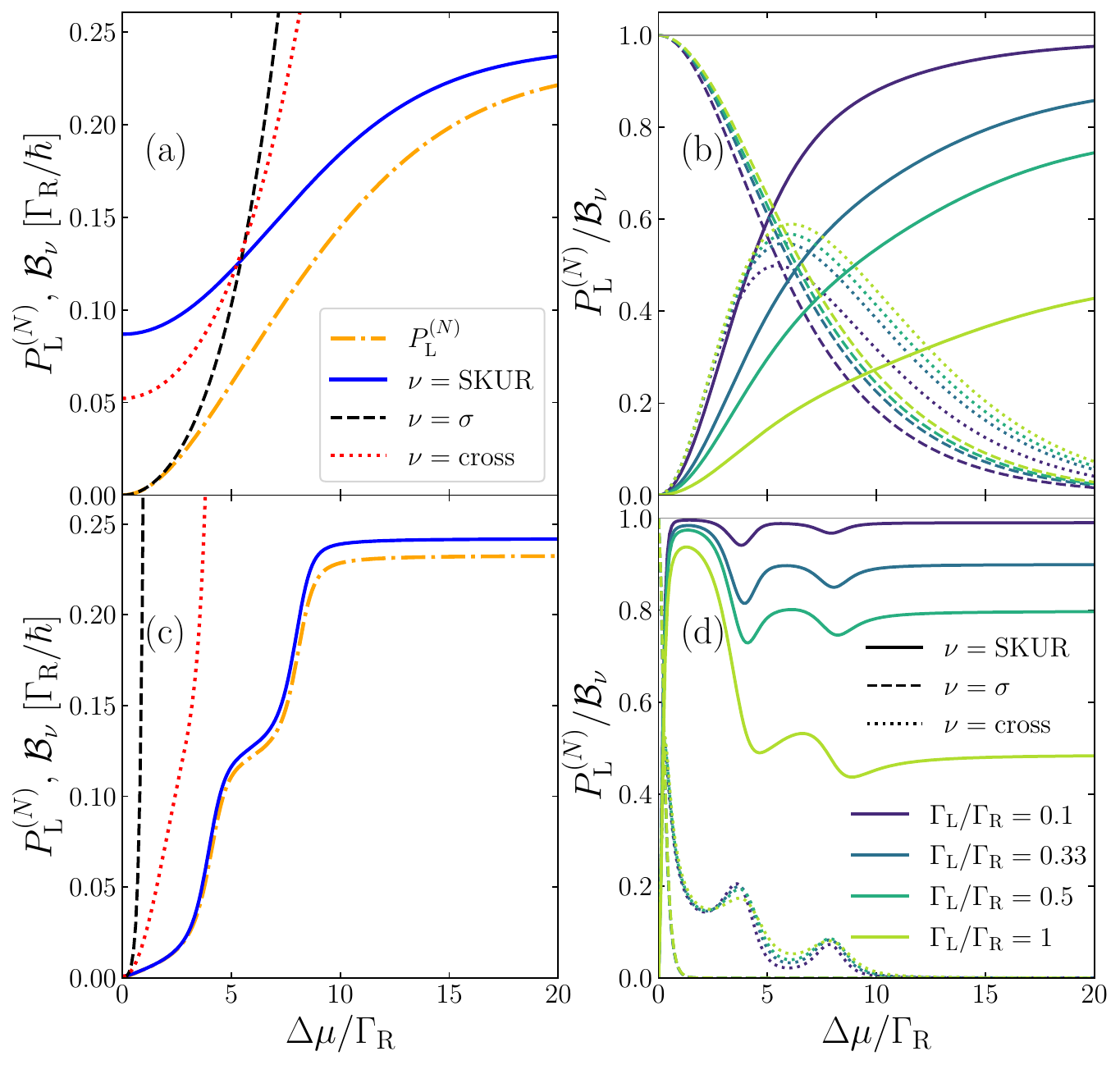}
\caption{{Comparison of the S-KUR~\eqref{eq: SKUR P bound SS}, with the precision bounds of~\eqref{eq: QTUR} and~\eqref{eq: QKUR} for particle transport through a DQD. In panels (a,c) we plot the particle-current precision and its upper bounds as functions of chemical potential bias $\Delta \mu$, entering $\mu_\MRL =\Delta\mu/2$ and $\mu_\MRR = -\Delta \mu/2$. In panels (b,d) we plot the ratio between precision and cost function of the bounds as functions of $\Delta \mu$ for multiple values of the coupling asymmetry $\Gamma_\MRL/\Gamma_\MRR$. In panels~(a,b) we use $\kB T_\MRL=\kB T_\MRR =2\Gamma_\MRR$ and in panels~(c,d) we use $\kB T_\MRL=\kB T_\MRR =\Gamma_\MRR/10$. In panels~(a,c) we use $\Gamma_\MRL = \Gamma_\MRR/3$ and for all panels we use $g = 3\Gamma_\MRR$ and $\epsilon_d = \Gamma_\MRR$.   }}
    \label{fig: comparision bounds}
\end{figure}
In panels~(a,c) of Fig.~\ref{fig: comparision bounds}, we show the particle-current precision and its upper bounds as function of the chemical-potential bias for two different temperature regimes. In panels~(b,d), we show how tight the different bounds are by plotting the precision divided by the different cost functions for different $\Gamma_\MRL/\Gamma_\MRR$. It is clear that quantum TUR of~\eqref{eq: QTUR} is the tightest bound close to equilibrium while the S-KUR is the tightest far away from equilibrium. For the case of $T_\MRL=T_\MRR =2\Gamma_\MRR$ in panels~(a,b) there is an intermediate bias regime where the bound of~\eqref{eq: QKUR} is tighter than the both the \SKUR\ and the quantum TUR of~\eqref{eq: QTUR} for symmetric and moderately asymmetric coupling. The S-KUR performs particularly well in the colder regime of $T_\MRL=T_\MRR =\Gamma_\MRR/10$ displayed in panels~(c,d) where both the bounds of Eqs.~\eqref{eq: QTUR} and \eqref{eq: QKUR} are loose far from equilibrium.

While the precision bound~\eqref{eq: QKUR} of Ref.~\cite{BlasiMay2025} can be saturated far from equilibrium this is not a generic feature. 
This can be understood when writing the cost function in terms of the thermal noise, $\mathcal{B}_\mathrm{cross}=(\mathscr{K}^\mathrm{cross}_\alpha) ^2/ S^{(N)\mathrm{th}}_\alpha$. The thermal noise $S^{(N)\mathrm{th}}_\MRL$ approaches zero when either the temperatures are low, or when a large symmetric chemical potential bias is applied, causing the reservoirs' Fermi levels to move away from the DQD's transmission window. This in turn causes the upper bound in~\eqref{eq: QKUR} to diverge. However, the precision remains finite whenever the shot noise gives an appreciable contribution to the overall noise, as would be the case for a large $\Delta\mu$ applied to a coherent conductor which is not fully transparent for all relevant energies. Hence, while the bound of~\eqref{eq: QKUR} can outperform the S-KUR in ideal settings, such as for perfect boxcar transmissions with zero shot noise, we expect the S-KUR to be a tighter bound when shot noise can not be neglected.

{\section{S-KUR applied to a finite-size system with interactions}\label{sec: DQD finite leads}}

To show the generality of the \SKUR\ we here present results for a complementary setting, namely \textit{transient} particle transfer through an \textit{interacting} double quantum dot coupled to two \textit{finite} leads. Similar models have been used to study finite-size effects in real-time simulations of transport~\cite{Al-Hassanieh2006May,daSilva2008Nov,Heidrich-Meisner2009Feb,Pires2020Mar}. Crucially, transport KURs for quantum systems developed previously for noninteracting or weakly-coupled systems~\cite{Palmqvist2025Oct,Prech2023Jun,Palmqvist2025Jul,BlasiMay2025,Hasegawa2020Jul,Tesser2026Jul} can not be applied to this finite-lead dynamical setting.

\subsection{Double quantum dot coupled to finite leads: Model and transport observables}

The total lead-DQD system evolves unitarily, so no weak-coupling or Markov approximation is made. Each lead $\alpha \in \{\MRL,\MRR\}$ is described by a tight-binding Hamiltonian with $W$ sites and open-boundary conditions~\cite{Datta1995Sep}
\begin{equation}
    \hat{H}_\alpha  =- \tau  \sum^{W-1}_{w=1} (\hat{c}^\dag_{\alpha,w} \hat{c}_{\alpha,w+1} +\hat{c}^\dag_{\alpha,w+1}\hat{c}_{\alpha,w} ).
\end{equation}
We diagonalize the isolated lead Hamiltonians by introducing the creation operator for orbital $q$
\begin{equation}
    \hat{a}^\dag_{\alpha,q} := \sum^{W}_{w=1} \phi_q(w) \hat{c}^\dag_{\alpha,w},
\end{equation}
where
\begin{equation}
    \phi_q(w) = \sqrt{\frac{2}{W+1}} \sin \left( \frac{\pi q w}{W+1}\right),
\end{equation}
such that
\begin{equation}
    \hat{H}_\alpha = \sum_{q=1}^W \epsilon_q \hat{a}^\dag_{\alpha,q} \hat{a}_{\alpha,q}, \, \epsilon_q := -2\tau  \cos\left( \frac{\pi q}{W+1}\right).
\end{equation}
The number operator of lead $\alpha$ is defined as $\hat{N}_\alpha =\sum^{W}_{w=1} \hat{c}^\dag_{\alpha,w} \hat{c}_{\alpha,w}$.
The Hamiltonian of the spinless, interacting DQD is given by
\begin{equation}
 \hat{H}_\mathrm{S} = \epsilon_d ( \hat{d}^\dag_1 \hat{d}_1 + \hat{d}^\dag_2 \hat{d}_2) +g (\hat{d}^\dag_1 \hat{d}_2 +\hat{d}^\dag_2 \hat{d}_1) +U\hat{d}^\dag_1 \hat{d}_1 \hat{d}^\dag_2 \hat{d}_2  .
\end{equation}
The couplings between the DQD and the leads read
\begin{equation}
\begin{aligned}
        \hat{V}_\MRL &= \tau_\MRL ( \hat{c}^\dag_{\MRL, W} \hat{d}_1 +  \hat{d}^\dag_1 \hat{c}_{\MRL, W}),\\
        \hat{V}_\MRR &= \tau_\MRR ( \hat{c}^\dag_{\MRR ,1} \hat{d}_2 +  \hat{d}^\dag_2 \hat{c}_{\MRR ,1} ).
\end{aligned}
\end{equation}
We initialize the system in a product state using a partitioned preparation~\cite{Ridley2018Jun} where the DQD is initially unoccupied, and the left and right leads contain $N_\MRL(0)$ and $N_\MRR(0)$ electrons initially,
\begin{equation}
    \ket{\psi(0)} = \left( \prod^{N_\MRL(0)}_{q=1} \hat{a}^\dag_{\MRL,q} \right) (\hat{d}^\dag_1)^0(\hat{d}^\dag_2)^0 \left( \prod^{N_\MRR(0)}_{q=1} \hat{a}^\dag_{\MRR,q} \right) \ket{0}.
\end{equation}
After the preparation, the total coupled system evolves unitarily.

 Since the initial state commutes with the number operator of the left lead\footnote{{In fact $\hat{N}_\MRL \ket{\psi(0)} = N_\MRL(0)\ket{\psi(0)} $.}}, we can apply the S-KUR to $\Delta \hat{N}_\MRL^\MRH(t) =  \hat{N}_\MRL^\MRH(t) - \hat{N}_\MRL^\MRH(0)$ while at the same time its mean $\langle\Delta \hat{N}_\MRL^\MRH(t)\rangle$ and variance $\mathrm{Var}[\Delta \hat{N}_\MRL^\MRH(t)]$ coincide with those of the TPM~\cite{Esposito2009Dec}. The particle-current operator evolves according to
\begin{equation}
    \hat{J}^{(N)}_\MRL (t) = \partial_t \Delta \hat{N}^{\MRH}_\MRL (t) = \frac{i}{\hbar} [\hat{V}^{\MRH}_\MRL(t), \hat{N}^{\MRH}_\MRL (t)],
\end{equation}
meaning that the relevant PDA is defined through the rescaling of the coupling strength to the left lead, $\hat{H}_\theta = \hat{H}_\MRS +\hat{H}_\MRL + \hat{H}_\MRR +\theta \hat{V}_\MRL +\hat{V}_\MRR$. Since the initial state is pure and the dynamics are unitary, the PDA coincides with the limiting activity 
\begin{equation}
    \mathscr{A}_\MRL(t) = \frac{1}{\hbar^2} \int_0^t d\tau \int_0^t d\tau' \langle\delta\hat{V}^\MRH_\MRL(\tau) \delta\hat{V}^\MRH_\MRL(\tau')\rangle. 
\end{equation}
To evaluate the PDA numerically, we propagate the derivative of the state, 
\begin{equation}
    \ket{\chi(t)} := \partial_\theta \ket{\psi(t,\theta)}\big|_{\theta=1},
\end{equation}
which evolves according to
\begin{equation}
    i\hbar \partial_t \ket{\chi (t)} = \hat{H} \ket{\chi (t)} + \hat{V}_\MRL \ket{\psi (t)},
\end{equation}
with initial condition $\ket{\chi(0)} = 0$. Note that $\ket{\chi(t)}$ does not represent a normalized state. From this, the PDA is calculated as~\cite{Braunstein1994May}
\begin{equation}
    \mathscr{A}_\MRL(t) =\langle\chi(t) |\chi(t)\rangle - |\langle\psi(t)|\chi(t)\rangle|^2.
\end{equation}
In addition, the response of the particle number is \begin{equation}
    \partial_\theta \langle \hat{N}^\MRH_\MRL (t,\theta)\rangle\big|_{\theta=1} =2 \Re \bra{\chi(t)} \hat{N}_\MRL\ket{\psi(t)}.
\end{equation}
Since the initial particle number is sharp, $\hat{N}_\MRL\ket{\psi(0)}=N_\MRL(0)\ket{\psi(0)}$, the mean and the variance of the transferred particle number are 
\begin{equation}
\begin{aligned}
        \langle\Delta \hat{N}_\MRL^\MRH(t)\rangle &= \bra{\psi(t)} \hat{N}_\MRL \ket{\psi(t)} -N_\MRL(0) ,\\
        \mathrm{Var}[\Delta \hat{N}_\MRL^\MRH(t)] &= \bra{\psi(t)} \hat{N}^2_\MRL \ket{\psi(t)} -\bra{\psi(t)} \hat{N}_\MRL \ket{\psi(t)}^2 .
\end{aligned}
\end{equation}

\begin{figure}
    \centering
   \includegraphics[width=3.3in]{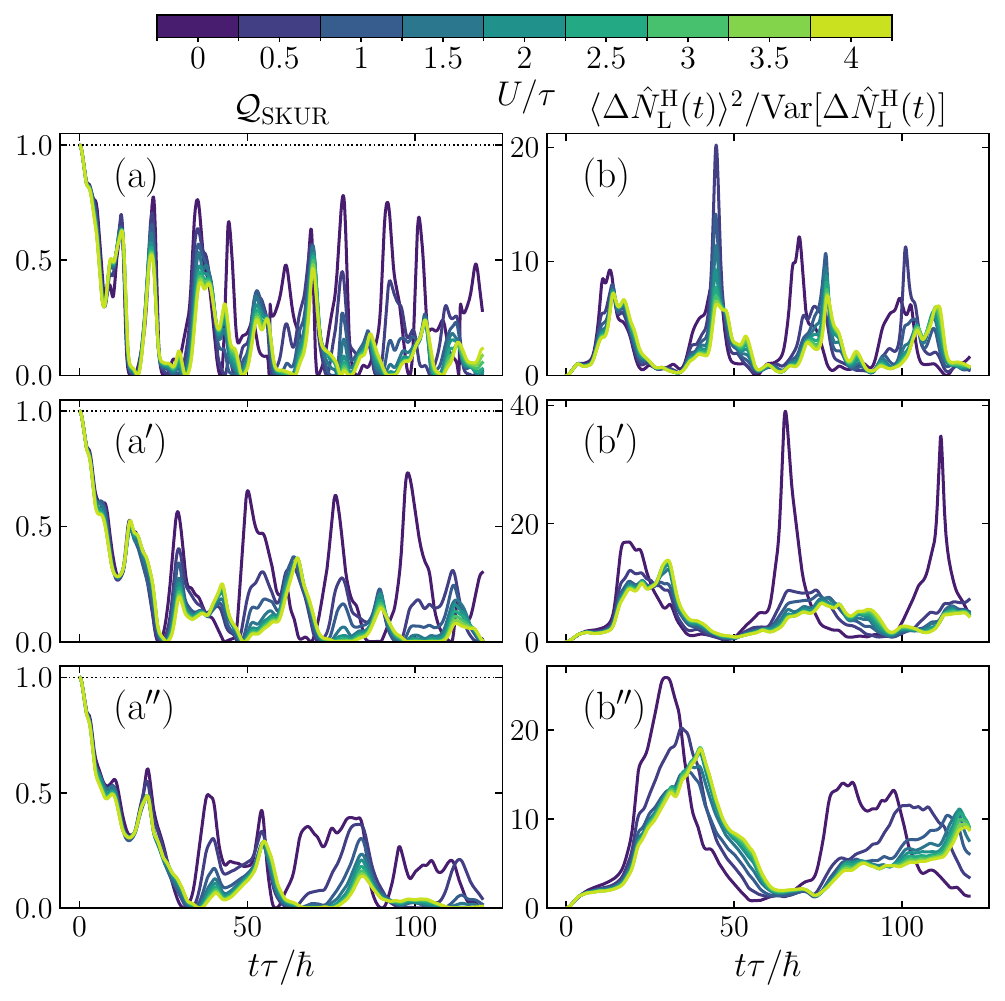}
\caption{{(a,a$'$,a$''$) Tightness of S-KUR. (b,b$'$,b$''$) Precision of particle transfers in interacting, finite-size system as functions of time. The first, second, and third rows show leads with $W=3,6,9$ sites, respectively. Different color lines show different strengths of the Coulomb interaction $U$. All panels use $\tau_\MRL=\tau_\MRR=\tau/2$, $g=\tau$ and $\epsilon_d =0$.}}
    \label{fig: finite size SKUR}
\end{figure}

\subsection{Predictiveness of the S-KUR for finite interacting systems}

In Fig.~\ref{fig: finite size SKUR} we demonstrate how tight the \SKUR\, is, quantified by the ratio
\begin{equation}\label{eq: finite time SKUR ratio}
    \mathcal{Q}_\mathrm{SKUR} = \frac{ \langle \Delta \hat{N}_\MRL^\MRH(t)\rangle ^2/\text{Var}[\Delta \hat{N}_\MRL^\MRH(t)] }{4  \mathscr{A}_\MRL(t)/\left( 1 +\Phi_\MRL^{(N)}(t) \right)^2} \leq 1.
\end{equation}
Together with this ratio, we show the precision in particle transfers as functions of time. We show results for different lead sizes, $W=3,6,9$, as different rows in the figure, where differently colored lines indicate different values of the Coulomb interaction $U$. In all panels, we use $\tau_\MRL=\tau_\MRR=\tau/2$ and $\epsilon_d=0$. We initialize the left lead with $N_\MRL(0) =  2W/3$ and the right lead with $N_\MRR(0)  = W/3$ particles, with the filling imbalance acting as a nonequilibrium condition. 

Interestingly, both the tightness of the \SKUR\ and the transfer precision display strong nonmonotonicity in time, with in particular $\mathcal{Q}_\mathrm{SKUR}$ displaying multiple revivals. These revivals are especially pronounced for the shorter lead lengths~\cite{Al-Hassanieh2006May,Pires2020Mar} and weak interaction. In particular, local minima in $\mathcal{Q}_\mathrm{SKUR}$ occur when $\partial_\theta \langle \hat{N}_\MRL^\MRH(t,\theta)\rangle|_{\theta=1}\approx0$, meaning that the particle number fails to capture the, in general, complicated dependence of the state on the coupling strength.
This dependence increases over time, as quantified by the partial dynamical activity, see Fig.~\ref{fig:app: SKUR finite size}. However, due to the finite size, the particle number of the left lead is able to regain the sensitivity to the coupling strength, leading to the revivals of $\mathcal{Q}_\mathrm{SKUR}$.

These results confirm the validity of the \SKUR\ for finite-time interacting dynamics, as expected, and show that the degree of tightness undergoes coherent decrease and revival. Additional results, displaying the detailed behavior of the PDA, the particle response, the mean particle transfer, and its variance underlying the discussed features, are discussed in Appendix~\ref{app: SKUR finite size}, Fig.~\ref{fig:app: SKUR finite size}.

\section{Conclusions}

We have established a generally valid susceptibility-kinetic uncertainty relation for quantum systems (\SKUR)~\eqref{eq:SKUR}, bounding the precision of arbitrary transport observables by a partial dynamical activity in Eq.~\eqref{eq:A_V}.
Our result provides the long-sought extension of the KUR beyond the classical or weakly-coupled regimes: The \SKUR\ is valid for generic currents, for finite-size systems and for systems with strong system-reservoir coupling, and in the presence of coherence,  particle interactions, and non-Markovian dynamics, thereby providing a universal precision bound.
In our approach, the partial and its limiting dynamical activity are derived from a microscopic picture by rescaling the system-reservoir coupling strength.
This information-geometric derivation unifies previously introduced notions of quantum dynamical activity: It reduces to the usual jump activity in the GKSL dynamics~\cite{Prech2025Jan}; it generalizes notions of activity based on rescaling of time~\cite{Hasegawa2023May, Nakajima2023Nov,Nishiyama2024Apr, Nishiyama2025Apr}; and it connects naturally to measurement-based notions of activity under weak coupling~\cite{Hasegawa2025Nov,VanVu2026May}.
The susceptibility $\dot{M}^{(X)}$ entering the \SKUR\ accounts for how the prior evolution responds to a change in the coupling strength. This response term guarantees the validity of the \SKUR\ in the presence of quantum effects and is experimentally accessible whenever the system-reservoir coupling can be tuned.
We demonstrate the validity and predictiveness of the \SKUR\ for steady-state transport by applying it to the experimentally relevant platform of charge- and energy transport through a double quantum dot. In this coherent transport setting, local classical KURs are strongly violated, while the \SKUR\ remains valid through the inclusion of $\dot{M}^{(X)}$ and is even tight in broad parameter regimes.

Our results offer a general framework for optimizing current precision in quantum (thermal) devices, ideal to study precision limits of realistic quantum devices where effects such as particle interactions, strong coupling, time-dependent driving, non-Markovianity, and finite-size reservoirs play key roles. The S-KUR can also serve as an inference tool for estimating the quantum dynamical activities in experiments where the system-reservoir coupling is tunable. Indeed, the inference of activity or ``traffic" can be of interest for chemical reactions~\cite{Aslyamov2025Apr}; recent works have also indicated that dynamical activity can play a key role in thermodynamic performance, when exploiting collective phenomena~\cite{Yunoki2026Apr,Toma2026Jul} and is therefore a quantity to be explored.

\section*{Acknowledgments}
We thank Juliette Monsel, Elsa Danielsson, Sofia Sevitz, Mark Mitchison, Nicolás Torres Domínguez,  and Gabriel Landi for helpful discussions and for providing useful comments on our work. We gratefully acknowledge funding from the European Research Council (ERC) under the European Union’s Horizon Europe research and innovation program (101088169/NanoRecycle).

\section*{Author contributions}
D.P. performed the majority of derivations as well as analytical and numerical calculations. J.S. contributed to discussions on the derivation of the S-KUR and to the analysis of results. L.T. contributed to derivations and result analysis and oversaw the project progress. All authors co-wrote the manuscript.

\appendix
\section{Overview of key quantities}\label{app:overview}
We collect an overview of all key quantities in Table~\ref{tab:overview}.

\begin{table}[h]
    \centering
    {\begin{tabular}{l|p{4.4cm}|p{1cm}}
       Quantity & Meaning & Eq.  \\
         \hline\hline
         $\hat{J}^{(X)}(t)$ & \raggedright current operator of $\hat{X}$, generated by $\hat{V}(t)$ & \eqref{eq:current}\\
         \hline
         $\Delta\hat{X}^\MRH(t)$ & \raggedright change of the observable $\hat{X}$ & \eqref{eq:DeltaX}\\
         \hline
         $\mathscr{A}_V(t)$ & \raggedright partial dynamical activity (PDA), $\mathcal{F}_{\theta\theta}/4$ & \eqref{eq:A_V} \\
         \hline
         $\mathscr{A}^\mathrm{lim}_V(t)$ & \raggedright limiting activity, upper bound on the PDA & \eqref{eq:Alim} \\
         \hline
         $\dot{M}^{(X)}(t)$ & \raggedright susceptibility, response of the current to the strength of $\hat{V}(t)$ & \eqref{eq:M-term} \\
         \hline
         $M^{(X)}(t)$ & \raggedright time-integrated susceptibility & \eqref{eq:SKUR-Phi} \\
         \hline
         $\mathcal{J}^{(X)}(t)$ & \raggedright response current, $\tfrac12\big(\langle\hat{J}^{(X)}\rangle+\dot{M}^{(X)}\big)$ & \eqref{eq: resp current} \\
         \hline
         $\Phi^{(X)}(t)$ & \raggedright quantum correction, $M^{(X)}/\langle\Delta\hat{X}^\MRH\rangle$ & \eqref{eq:SKUR-Phi} \\
         \hline
         $S^{(X)}$ & \raggedright zero-frequency noise of $\hat{J}^{(X)}$ & \eqref{eq:noise} \\
         \hline
         $\mathscr{K}_V,\ \mathscr{K}^\mathrm{lim}_V$ & \raggedright steady-state rates of the PDA and of the limiting activity & \eqref{eq:SKUR-ss}, \eqref{eq:Klim} \\
         \hline
         $\mathscr{P}^{(X)}_\alpha$ & \raggedright precision of the response current (\mbox{S-KUR}) & \eqref{eq: ss transport relax MKUR} \\
         \hline
         $P^{(X)}_\alpha$ & \raggedright precision of the current  & \eqref{eq:classical-KUR} \\
         \hline
         $\mathcal{Q}_\mathrm{SKUR}$ & \raggedright tightness ratio of the \mbox{S-KUR}, $\leq 1$ & \eqref{eq: finite time SKUR ratio} \\
         \hline
    \end{tabular}}
    \caption{{Overview of key quantities, their meaning, and the equations in which they are defined.}}
    \label{tab:overview}
\end{table}

\section{Quantum Fisher Information and response}\label{app:Quantum-Fisher-Information}
In this appendix, we give a brief introduction to the Quantum Fisher information (QFI). For two density operators \(\hat{\rho}_1\) and \(\hat{\rho}_2\), the Bures distance is defined by
\begin{equation}
   \begin{aligned}
        D^2_\mathrm{B}(\hat{\rho}_1,\hat{\rho}_2) &= 2-2\sqrt{\mathrm{Fid}(\hat{\rho}_1,\hat{\rho}_2)},\\ \mathrm{Fid}(\hat{\rho}_1,\hat{\rho}_2)&= \left(\tr{\sqrt{\sqrt{\hat{\rho}_1}\hat{\rho}_2\sqrt{\hat{\rho}_1}}} \right)^2.
   \end{aligned}
\end{equation}
Consider a state $\hat{\rho}(\theta)$ smoothly parametrized by a real number $\theta$ with fixed rank. The QFI $\mathcal{F}_{\theta\theta}$ is proportional to the Bures metric, and gives the infinitesimal Bures distance 
\begin{equation}
    D_{\rm B}^2(\hat{\rho}(\theta),\hat{\rho}(\theta+d\theta))=\frac{1}{4}\mathcal{F}_{\theta\theta}\,d\theta^2 .
\end{equation}
The QFI can be expressed using the symmetric logarithmic derivative $\hat{\mathfrak{L}}(\theta)$, as~\cite{Helstrom1969Jun,Holevo}
\begin{equation}\label{eq: def QFI}
\begin{split}
    \mathcal{F}_{\theta\theta}&= \tr{\hat{\rho}(\theta) \hat{\mathfrak{L}}^2(\theta)}, \\
    \partial_{\theta}  \hat{\rho}(\theta)&= \frac{1}{2} \big(\hat{\rho}(\theta)\hat{\mathfrak{L}}(\theta) +\hat{\mathfrak{L}}(\theta) \hat{\rho}(\theta)\big).
    \end{split}
\end{equation}
Importantly, the QFI enters the quantum Cram\'er-Rao bound, where for an observable $\hat{X}$
\begin{equation}
    \frac{(\partial_\theta \mathrm{tr}\{ \hat{X} \hat{\rho}(\theta) \} )^2 }{ \mathrm{tr}\{ (\hat{X} -\mathrm{tr}\{ \hat{X} \hat{\rho}(\theta) \})^2 \hat{\rho}(\theta) \} } \leq \mathcal{F}_{\theta\theta}.
\end{equation}

Using the spectral decomposition of a density matrix $\hat{\rho}(\theta) = \sum^{d-1}_{i=0} p_i(\theta) \ket{\psi_i(\theta)}\bra{\psi_i(\theta)}$ of rank $d$, the QFI in Eq.~\eqref{eq: def QFI} can be expressed as~\cite{Liu2019Dec}
\begin{equation}\label{eq: app QFI mix}
    \mathcal{F}_{\theta\theta}= \sum^{d-1}_{\substack{i,j=0,\\ p_i+p_j\neq 0}} \frac{2|\bra{\psi_i (\theta)} (\partial_\theta \hat{\rho}(\theta))  \ket{\psi_j (\theta)} |^2  }{p_i(\theta) +p_j(\theta)}.
\end{equation}
If the state is pure $\hat{\rho}(\theta) = \ket{\psi(\theta)}\bra{\psi(\theta)}$, the situation is simplified as the Bures metric reduces to the Fubini-Study metric and the QFI becomes
\begin{equation}
    \begin{aligned}
         \mathcal{F}_{\theta\theta} = 4 \Re [\langle \partial_\theta \psi(\theta)|\partial_\theta \psi(\theta)\rangle - |\langle \partial_\theta \psi(\theta)| \psi(\theta)\rangle|^2 ].
    \end{aligned}
\end{equation}

In the case of the PDA, the time evolution under the deformed Hamiltonian $\hat{H}_\theta(t) = \hat{H}_0(t) + \theta \hat{V}(t)$ results in the parametrization of the state. When the initial state is pure $\hat{\rho}(0) =\ket{\psi(0)} \bra{\psi(0)}$, the parametrized state and its variation are 
\begin{equation}
\begin{split}
    \ket{\psi(t, \theta)} &:= \hat{U}_\theta(t,0) \ket{\psi(0)},\\
    \ket{\partial_\theta \psi(t, \theta)} &:= \partial_\theta \hat{U}_\theta(t,0) \ket{\psi(0)}.
\end{split}
\end{equation}
The derivative of the unitary is computed as
\begin{equation}\label{eq: deriv U}
    \begin{aligned}
        \partial_\theta \hat{U}_\theta(t,0) &= -\frac{i}{\hbar}\hat{U}_\theta(t,0)\int_0^t ds \hat{U}_\theta^\dag(s,0) ({\partial_\theta \hat{H}_\theta(s)}) \hat{U}_\theta(s,0)\\ &= -\frac{i}{\hbar}\hat{U}_\theta(t,0) \int_0^t ds \hat{V}^\mathrm{H}(s,\theta),
    \end{aligned}
\end{equation}
where $\hat{V}^\mathrm{H} (s,\theta)= \hat{U}_\theta^\dag(s,0) \hat{V}(s) \hat{U}_\theta(s,0)$ is the operator in the Heisenberg picture, evolving under the $\theta$-modified dynamics. Inserting Eq.~\eqref{eq: deriv U} into the expressions of the QFI results in the partial dynamical activity of Eq.~\eqref{eq:A_V} used in the main text.

We show here that the responses of $\langle\Delta\hat{X}^\MRH(t,\theta)\rangle$ and $\tr{\hat{Y}\hat{\rho}(t,\theta)}$ coincide. Using Eq.~\eqref{eq: deriv U}, we find
\begin{equation}
   \begin{aligned}
        \partial_\theta \tr{ \hat{Y} \hat{\rho}(t,\theta) }\Big|_{\theta=1} &= \partial_\theta \tr{\hat{X}^\MRH(t,\theta) \hat{\rho}(0)}\Big|_{\theta=1} \\ &-\frac{i}{\hbar} \int^t_0ds \tr{ \hat{\rho}(0) [ \hat{V}^\MRH (s), \hat{X}]}.
   \end{aligned}
\end{equation}
By the cyclic property of the trace, $\tr{ \hat{\rho}(0) [ \hat{V}^\MRH (s), \hat{X}]} = \tr{[\hat{X},\hat{\rho}(0)]\hat{V}^\MRH (s)}$, which vanishes for $[\hat{\rho}(0),\hat{X}]=0$.

\section{Weak-coupling limit of $\dot{M}^{(X)}$}\label{app:weak-coupling-M}
We show here that when the coupling $\hat{V}(t)$ can be treated perturbatively, the average current $J^{(X)}$ and the susceptibility $\dot{M}^{(X)}$ coincide, i.e. $J^{(X)}=\dot{M}^{(X)} = \mathcal{J}^{(X)}$.

The unitary evolution $\hat{U}_0(t,0)$ in the absence of coupling is given by
\begin{equation}
    \hat{U}_0(t,0):= T_+\exp\left\{-\frac{i}{\hbar}\int_0^t \hat{H}_0(s)ds\right\}
\end{equation}
and $\hat{V}^\text{I}(s):= [\hat{U}_0(s,0)]^\dagger \hat{V}(s) \hat{U}_0(s,0)$ is the coupling in the interaction picture.
Then, we expand $J^{(X)}$ up to second order in the coupling, namely $J^{(X)} \approx J^{(X)}_1 + J^{(X)}_2$ with
\begin{subequations}
    \begin{align}
        J^{(X)}_1 &= \frac{i}{\hbar}\Tr\left\{[\hat{V}^\text{I}(t), \hat{X}] \hat{\rho}_0\right\},\\
        J^{(X)}_2 &= \frac{1}{\hbar^2}\int_0^t ds \Tr\left\{[[\hat{V}^\text{I}(t), \hat{X}], \hat{V}^\text{I}(s)]\hat{\rho}_0\right\},\label{app:eq:J_2}
    \end{align}
\end{subequations}
where we used that $[\hat{X}, \hat{H}_0(t)]=0\,\forall t$. Furthermore, when $[\hat{X}, \hat{\rho}_0]=0$, the first order contribution to the current vanishes, $J^{(X)}_1=0$, and the second order $J^{(X)}_2$ becomes the leading contribution.

For the susceptibility $\dot{M}^{(X)}$, we use Eq.~\eqref{eq: deriv U} at $\theta = 1$ to find
\begin{equation}\label{app:eq:M}
    \dot{M}^{(X)} = \frac{1}{\hbar^2}\int_0^t ds \langle [[\hat{V}^\MRH(t), \hat{X}^\MRH(t)], \hat{V}^\MRH(s)]\rangle.
\end{equation}
At lowest order in $\hat{V}$, the leading order of the average current [Eq.~\eqref{app:eq:J_2}] and the susceptibility [Eq.~\eqref{app:eq:M}] coincide.

\section{$\mathscr{K}_\alpha^\text{lim}$ in GKSL dynamics}\label{app: GKSL klim}
In this section, we show that the limiting activity, Eq.~\eqref{eq:Alim}, reduces to the total number of jumps observed in the context of GKSL dynamics when the Lamb shift is neglected. We consider the Hamiltonian of a system coupled to reservoirs labeled by $\alpha$
\begin{equation}
    \hat{H} = \hat{H}_\MRS + \sum_{\alpha} (\hat{H}_\alpha + \hat{V}_\alpha),
\end{equation}
where the interaction between reservoir $\alpha$ and the central system is given by
\begin{equation}
    \hat{V}_\alpha = \sum_k \hat{A}_{\alpha k} \otimes \hat{B}_{\alpha k} = \sum_k \hat{A}^\dag_{\alpha k} \otimes \hat{B}_{\alpha k}^\dag.
\end{equation}
Here $\hat{A}_{\alpha k}$ acts on the system and $\hat{B}_{\alpha k}$ on reservoir $\alpha$. We study the corresponding limiting activity rate
\begin{equation}
\begin{aligned}
        \mathscr{K}^\mathrm{lim}_\alpha (t) = \frac{1}{\hbar^2} \int_0^tds\langle \{\delta\hat{V}_\alpha^\mathrm{H}(t), \delta\hat{V}_\alpha^\mathrm{H}(s)\} \rangle \\
         =\frac{2}{\hbar^2} \Re \int_0^tds\langle \delta\hat{V}_\alpha^\mathrm{H}(t) \delta\hat{V}_\alpha^\mathrm{H}(s) \rangle.
\end{aligned}
\end{equation}
Defining $\hat{H}_0 :=\hat{H}_\MRS + \sum_\alpha \hat{H}_\alpha $ and $\hat{V}:= \sum_\alpha \hat{V}_\alpha$, the time evolutions to switch to the interaction picture are
\begin{equation}
    \begin{aligned}
        \hat{U}_0(t,0) &= {T}_+\exp\left\{-\frac{i}{\hbar} \int^t_{0} d\tau \hat{H}_0(\tau) \right\} ,\\
        \hat{U}_\mathrm{I}(t,0) &= {T}_+\exp\left\{-\frac{i}{\hbar} \int^t_{0} d\tau \hat{V}^\mathrm{I}(\tau) \right\},
    \end{aligned}
\end{equation}
where $\hat{V}^\mathrm{I}(t)= \hat{U}^\dag_0(t,0)\hat{V}\hat{U}_0(t,0)$. The density operator in the interaction picture at time $t$ is given by $\hat{\rho}^\mathrm{I}(t) = \hat{U}_\mathrm{I}(t,0) \hat{\rho}(0) \hat{U}^\dag_\mathrm{I}(t,0)$. The expression for the activity used for GKSL dynamics is second order in the interaction and linear in the density matrix. Keeping terms which are at most second order in $\hat{V}^\mathrm{I}$, we get~\cite{Palmqvist2026}
\begin{equation}
    \begin{aligned}
        &\tr{\hat{V}_\alpha^\mathrm{H}(t) \hat{V}^\mathrm{H}_\alpha(s) \hat{\rho}(0) }  \approx \tr{\hat{V}_\alpha^\mathrm{I}(t) \hat{V}_\alpha^\mathrm{I}(s) \hat{\rho}^\mathrm{I}(t)}.
    \end{aligned}
\end{equation}
This truncation retains only correlations within the reservoir correlation time. We drop correlations over the (slower) system time scale, as they stem from the Lamb shift. They are included in Appendix~\ref{app:Lamb-shift}. 

Next, we assume that the spectrum of $\hat{H}_\MRS$ is discrete and denote the eigenvalues of the system Hamiltonian as $\epsilon$ with the projector onto the eigenspace of $\epsilon$ as $\hat{\Pi}(\epsilon)$. We introduce
\begin{equation}
    \hat{A}_{\alpha k}(\omega) =\sum_{\epsilon'-\epsilon=\omega} \hat{\Pi}(\epsilon) \hat{A}_{\alpha k} \hat{\Pi}(\epsilon'),
\end{equation}
with the properties~\cite{Breuer2007}
\begin{equation}
    [\hat{H}_\MRS, \hat{A}_{\alpha k}(\omega)]= -\hbar\omega \hat{A}_{\alpha k}(\omega),\,\hat{A}^\dag_{\alpha k}(\omega) = \hat{A}_{\alpha k}(-\omega). 
\end{equation}
Under the assumption that there is no explicit time dependence, the operators in the interaction picture evolve as
\begin{align}
    \hat{A}^\mathrm{I}_{\alpha k}(\omega;t)&=e^{i  \hat{H}_\MRS t/\hbar } \hat{A}_{\alpha k}(\omega) e^{-i  \hat{H}_\MRS t/\hbar } = e^{-i \omega t}\hat{A}_{\alpha k}(\omega),\\
    \hat{B}^\mathrm{I}_{\alpha k}(t)&=e^{i  \hat{H}_\alpha t/\hbar } \hat{B}_{\alpha k} e^{-i  \hat{H}_\alpha t/\hbar }.
\end{align}
Moreover, we assume $\langle \hat{B}^\mathrm{I}_\alpha(t)\rangle=0$ and weak coupling, such that we can use the Born approximation $\hat{\rho}^\mathrm{I}(t) \approx \hat{\rho}^\mathrm{I}_\MRS(t)\bigotimes_\alpha \hat{\rho}_\alpha$, where $\hat{\rho}_\alpha$ are density matrices of the reservoirs. We approximate
\begin{equation}
   \begin{aligned}
        &\mathscr{K}^\mathrm{lim}_\alpha (t)\approx \Re \sum_{kk'} \int^t_0\frac{2ds}{\hbar^2} \Big( \tr{\hat{A}^\mathrm{I}_{\alpha k'} (t) \hat{A}^\mathrm{I}_{\alpha k} (s) \hat{\rho}^\mathrm{I}_\MRS(t)} \times \\  &\quad\quad\quad\quad\quad\quad\times \tr{ \hat{B}^{\mathrm{I}}_{\alpha k'} (t) \hat{B}^{\mathrm{I}}_{\alpha k} (s) \hat{\rho}_\alpha} \Big)\\
    &\qquad=\Re \sum_{kk'}  \sum_{\omega\omega'} e^{-i(\omega  +\omega' )t} \tr{\hat{A}_{\alpha k'}(\omega') \hat{A}_{\alpha k}(\omega) \hat{\rho}^\mathrm{I}_\MRS(t)} \times\\
    &\quad\quad\quad\times \int^t_{0} \frac{2ds}{\hbar^2}e^{-i \omega' (s-t)} \tr{\hat{B}^{\mathrm{I} \dag}_{\alpha k'} (t) \hat{B}^{\mathrm{I}}_{\alpha k} (s) \hat{\rho}_\alpha}.\label{eq:Klim_approx}
   \end{aligned}
\end{equation}

Next, we use the Markov approximation where the reservoir correlation functions decay fast compared to the time scale of the system dynamics and extend the upper integration limit to $\infty$. We introduce the reservoir correlation functions
\begin{equation}
\begin{aligned}
        \Gamma^\alpha_{k'k}(\omega) &= \int^\infty_{0} \frac{ds}{\hbar^2}e^{i \omega s} \tr{\hat{B}^{\mathrm{I} \dag}_{\alpha k'} (s) \hat{B}^\mathrm{I}_{\alpha k} (0) \hat{\rho}_\alpha}\\
        &= \frac{1}{2}\gamma^{\alpha}_{k'k} (\omega) + i\Delta^\alpha_{k'k}(\omega)
\end{aligned}
\end{equation}
where
\begin{equation}
    \gamma^\alpha_{k'k}(\omega)=\int_{-\infty}^{\infty} \frac{ds}{\hbar^2}e^{i\omega s}\tr{\hat{B}_{\alpha k'}^{\mathrm{I} \dag} (s)\hat{B}_{\alpha k}^\mathrm{I}(0)\rho_\alpha}
\end{equation}
is the positive semidefinite rate matrix, while
$\Delta^\alpha(\omega)$ gives the principal-value contribution associated with the Lamb shift.

Next, we employ the secular approximation, neglecting any oscillating terms where $\omega +\omega' \neq 0$, yielding
\begin{equation}
   \begin{aligned}
        \mathscr{K}^\mathrm{lim}_\alpha(t) 
    &=2\Re  \sum_{kk'}\sum_{\omega}  W^\alpha_{k'k}(\omega)\Gamma^\alpha_{k'k}(\omega).
   \end{aligned}
\end{equation}
Here we introduced the Hermitian matrix $W^\alpha_{k'k} =\tr{\hat{A}^\dag_{\alpha k'}(\omega) \hat{A}_{\alpha k}(\omega) \hat{\rho}^\mathrm{I}_\MRS(t)} $. Since $\gamma^{\alpha}_{k'k}(\omega)$ and $\Delta^{\alpha}_{k'k}(\omega)$ are both Hermitian matrices, taking the real part in Eq.~\eqref{eq:Klim_approx} reduces the expression to
\begin{equation}
   \begin{aligned}
        \mathscr{K}^\mathrm{lim}_\alpha(t) 
    &=\Re  \sum_{\omega}  \tr{ [W^\alpha(\omega)]^T( \gamma^{\alpha} (\omega) + 2i\Delta^\alpha(\omega)) } \\
    &= \sum_{kk'}\sum_{\omega}  \tr{\hat{A}^\dag_{\alpha k'}(\omega) \hat{A}_{\alpha k}(\omega) \hat{\rho}^\mathrm{I}_\MRS(t)} \gamma^\alpha_{k'k}(\omega).
   \end{aligned}
\end{equation}
By diagonalizing $\gamma^\alpha_{k'k}(\omega)$, we write the limiting activity in terms of the standard jump operators
\begin{equation}\label{eq: Klim diag}
   \begin{aligned}
        \mathscr{K}^\mathrm{lim}_\alpha(t) = \sum_\omega\sum_j  \tr{\hat{L}^\dag_{\alpha j}(\omega) \hat{L}_{\alpha j}(\omega) \hat{\rho}^\mathrm{I}_\MRS(t)}.
   \end{aligned}
\end{equation}
The number of jumps induced by $\alpha$ is given by $\mathscr{A}^\mathrm{lim}_\alpha(t) = \int_0^t ds \mathscr{K}^\mathrm{lim}_\alpha(s)$ and the total number of jumps is obtained by summing over reservoirs $\sum_\alpha \mathscr{A}^\mathrm{lim}_\alpha(t)$.

\section{\SKUR\ for GKSL dynamics}\label{app: SKUR GKSL}
In this appendix, we show how the \SKUR\ is related to the quantum KUR of Ref.~\cite{Prech2025Jan} for GKSL dynamics.

The microscopic deformation $\hat{H}_\theta = \hat{H}_\mathrm{S} + \hat{H}_\mathrm{E} + \theta \hat{V},$ results in an effective $\theta^2$ rescaling of the dissipative part of the Liouvillian in Eq.~\eqref{eq: deformed GKSL 1}, given that we can neglect the Lamb shift.\footnote{{For situations where the Lamb shift can \textit{not} be neglected, see Appendix~\ref{app:Lamb-shift} below.}} Also, the current inherits this effective rescaling
\begin{equation}
    J^{(X)}(\theta) = \theta^2 \sum_j x_{j} \tr{\hat{L}^\dag_j \hat{L}_j \hat{\rho}^\mathrm{ss}(\theta) },
\end{equation}
where $\mathcal{L}_\theta \hat{\rho}^\mathrm{ss}(\theta) =0$. Here $x_j$ is the increment of $\hat{X}$ being transported due to jump $j$. We can split the response in $J(\theta)$ into two parts
\begin{equation}
\begin{aligned}
        \partial_\theta J^{(X)}(\theta)\big|_{\theta =1} &= 2 J^{(X)} +\dot m^{(X)}, \\
   \dot m^{(X)}&= \partial_\theta \left(\frac{J^{(X)}(\theta)}{\theta^2} \right)\Bigg|_{\theta =1}.
\end{aligned}
\end{equation}
Here, to account for the explicit rescaling of the current, we divide by $\theta^2$ in $\dot m^{(X)}$ which should be distinguished from $\dot M^{(X)}$ appearing in the \SKUR. To connect to the KUR derived in Ref. \cite{Prech2025Jan}, we note that 
\begin{equation}
\begin{aligned}
        \dot m^{(X)} &= 2\tr{ \mathcal{I} \mathcal{L}^+ \mathcal{H}_\text{S} \hat{\rho}^\mathrm{ss}} = 2J^{(X)} \psi, \\
        \psi &= \frac{\tr{ \mathcal{I} \mathcal{L}^+ \mathcal{H}_\text{S} \hat{\rho}^\mathrm{ss}}}{J^{(X)}},
\end{aligned}
\end{equation}
where $\mathcal{I} \bullet = \sum x_j \hat{L}_j \bullet \hat{L}^\dag_j$ is the current superoperator with weights $x_j$, $\mathcal{L}^+ $ is the Drazin inverse of $\mathcal{L}$, $ \mathcal{H}_\text{S} \bullet =-\frac{i}{\hbar} [ \hat{H}_\text{S}, \bullet]$, and $\psi$ is the quantum correction present in the $\psi$-KUR derived in Ref.~\cite{Prech2025Jan}. To relate the $\psi$-KUR to the \SKUR\ we notice that
\begin{equation}
    \dot M^{(X)} = \partial_\theta \left(\frac{J^{(X)}(\theta)}{\theta} \right) \Bigg|_{\theta=1} =  J^{(X)} + 2 \psi J^{(X)},
\end{equation}
with the response current
\begin{equation}\label{eq:Psi_meaning}
    \mathcal{J}^{(X)} = \frac{1}{2}(J^{(X)} + \dot M^{(X)}) = J^{(X)}(1+\psi).
\end{equation}
Hence, the \SKUR\ in the limit of steady-state GKSL dynamics is equivalent to the $\psi$-KUR derived in Ref.~\cite{Prech2025Jan}
\begin{equation}
    \frac{(J^{(X)} + \dot M^{(X)})^2}{4S^{(X)}} = \frac{(J^{(X)} )^2 (1+\psi)^2}{S^{(X)}} \leq \sum_j \tr{\hat{L}^\dag_j \hat{L}_j \hat{\rho}^\mathrm{ss}}.
\end{equation}
Moreover, Eq.~\eqref{eq:Psi_meaning} provides an operational meaning to the quantum correction $\psi$ by relating it to the current response to a change in the system-environment coupling strength. When the steady-state density matrix commutes with the system Hamiltonian $ [ \hat{H}_\text{S}, \hat{\rho}]=0$, the dynamics are effectively classical and $\psi =0$~\cite{Landi2024Apr}, and as expected $\dot{M}^{(X)} =J^{(X)}$, meaning that the \SKUR\ reduces to the standard classical KUR.

This equivalence of the \SKUR\ and the $\psi$-KUR follows when the deformation of the underlying dynamics results in a rescaling of the full dissipative part of the dynamics. The local \SKUR\
\begin{equation}\label{eq: skur gksl}
    \frac{(J_\alpha^{(X)} + \dot M^{(X)}_\alpha)^2}{4S_\alpha^{(X)}} \leq \sum_j \tr{\hat{L}^\dag_{\alpha j} \hat{L}_{\alpha j} \hat{\rho}^\mathrm{ss}}.
\end{equation}
for transport is more refined: one may rescale the coupling strength to a single reservoir $\alpha$, resulting in only the dissipator associated with the same reservoir being rescaled in the Liouvillian. As a consequence, only the activity with respect to $\alpha$ enters the \SKUR\ and in general $\dot M^{(X)}_\alpha\neq J_\alpha^{(X)}$ even for classical dynamics. 
Thus, the \SKUR\ provides even an improvement of the classical KUR since it, in some situations, results in a tighter bound than the original KUR by avoiding the need to use a total activity.

\section{Lamb shift parametrization in the GKSL equation}\label{app:Lamb-shift}

In general, the microscopic parametrization $\hat{H}_\theta = \hat{H}_0 + \theta\hat{V}$ leads to a rescaling of not only the dissipator, but also of the Lamb-shift Hamiltonian~\cite{Breuer2007}. 
Consequently, the parametrized Liouvillian reads ($\hbar=1$)
\begin{equation}
\begin{split}
    &\partial_t \hat{\rho}_\text{S}(t, \theta) = \mathcal{L}_\theta \hat{\rho}_\text{S}(t,\theta)\\
    &=-i[\hat{H}_\text{S} + \theta^2 \hat{H}_\text{LS}, \hat{\rho}_\text{S}] + \theta^2 \sum_{k}\gamma_{k}\mathcal{D}[\hat{\ell}_{k}]\hat{\rho}_\text{S},
\end{split}
\end{equation}
where $\hat{H}_\text{S}$ is the bare Hamiltonian of the system, and $\hat{H}_\text{LS}$ is the Lamb shift contribution:
\begin{equation}
    \hat{H}_\text{LS} = \sum_{k} \Delta_{k}\hat{\ell}_{k}^\dagger\hat{\ell}_{k},
\end{equation}
where, for simplicity, we assume that the Lamb-shift and rate matrices are diagonal in the same basis.
Note that the operators $\hat{\ell}_k$ are related to the jump operators $\hat{L}_k$ in the main text through $\hat{L}_k = \sqrt{\gamma_k}\hat{\ell}_k$.

In the long-time limit, the quantum Fisher information {of the output state} is obtained following Ref.~\cite{Gammelmark2014Apr, Prech2025Jan}
\begin{equation}\label{app:eq:LS-QFI}
\begin{split}
    \mathcal{F}_{\theta\theta} &= 4t\sum_{k}\Tr\left\{\gamma_k(\partial_\theta \hat{\ell}_{k,\theta})^\dagger(\partial_\theta \hat{\ell}_{k,\theta})\hat{\rho}^\text{ss}\right\}+\\
    &-4t\Tr\left\{\mathcal{L}_\text{L}\mathcal{L}^+\mathcal{L}_\text{R}\hat{\rho}^\text{ss} +\mathcal{L}_\text{R}\mathcal{L}^+\mathcal{L}_\text{L}\hat{\rho}^\text{ss} \right\}
\end{split}
\end{equation}
where
\begin{equation}
    \begin{split}
        \mathcal{L}_\text{L}\hat{\rho} &:= -i\partial_\theta\hat{H}_{\mathrm{S},\theta}\hat{\rho} - \frac{1}{2}\sum_k \gamma_k \partial_\theta (\hat{\ell}_{k,\theta}^\dagger \hat{\ell}_{k,\theta})\hat{\rho}+ \\
        &\qquad+ \sum_k \gamma_k(\partial_\theta \hat{\ell}_{k,\theta})\hat{\rho}\hat{\ell}^\dagger_{k,\theta}, \\
        \mathcal{L}_\text{R}\hat{\rho} &:= i\hat{\rho}\partial_\theta\hat{H}_{\mathrm{S},\theta} - \frac{1}{2}\sum_k \gamma_k \hat{\rho}\partial_\theta (\hat{\ell}_{k,\theta}^\dagger \hat{\ell}_{k,\theta})+\\
        &\qquad+ \sum_k \gamma_k \hat{\ell}_{k,\theta}\hat{\rho}(\partial_\theta\hat{\ell}_{k,\theta})^\dagger. \\
    \end{split}
\end{equation}
The parametrization consists of $\hat{\ell}_{k,\theta} = \theta \hat{\ell}_k$ and $\hat{H}_{\mathrm{S},\theta} =\hat{H}_\text{S} + \theta^2 \hat{H}_\text{LS}$, such that $\mathcal{L}_\text{L/R}$ become
\begin{equation}\label{app:eq:LS-contributions}
    \begin{split}
        \mathcal{L}_\text{L}\hat{\rho} &= -2\theta\sum_k\left(i\Delta_k+\frac{\gamma_k}{2}\right)\hat{\ell}_k^\dagger\hat{\ell}_k\hat{\rho} + \theta \sum_k\gamma_k \hat{\ell}_k\hat{\rho}\hat{\ell}_k^\dagger, \\
        \mathcal{L}_\text{R}\hat{\rho} &= 2\theta\sum_k\left(i\Delta_k - \frac{\gamma_k}{2}\right)\hat{\rho}\hat{\ell}_k^\dagger \hat{\ell}_k + \theta\sum_k \gamma_k \hat{\ell}_k\hat{\rho}\hat{\ell}_k^\dagger. \\
    \end{split}
\end{equation}
The contributions in Eq.~\eqref{app:eq:LS-contributions} proportional to $\gamma_k$ are traceless, such that the second term in the quantum Fisher information in Eq.~\eqref{app:eq:LS-QFI} reads
\begin{equation}\label{app:eq:LS-QFI-extra}
    \begin{split}
        &-4t\Tr\left\{2i\theta \sum_k\Delta_k \hat{\ell}_k^\dagger\hat{\ell}_k\,
        \mathcal{L}^+(\mathcal{L}_\text{L} - \mathcal{L}_\text{R})\hat{\rho}^\text{ss}\right\} =\\
        &= -4t\Tr\left\{4i\theta^2 \sum_k\Delta_k \hat{\ell}_k^\dagger\hat{\ell}_k\,
        \mathcal{L}^+\cdot\right.\\
        &\qquad\cdot\left.\sum_q \left(-i\Delta_q\{\hat{\ell}_q^\dagger\hat{\ell}_q,\hat{\rho}^\text{ss}\}
        -\frac{\gamma_q}{2}[\hat{\ell}_q^\dagger\hat{\ell}_q,\hat{\rho}^\text{ss}]\right)\right\} \\
        &= -8t\theta^2 \Tr\left\{\mathscr{H}_\text{eff}\,\mathcal{L}^+\mathscr{H}_\text{eff}\,
        \hat{\rho}^\text{ss}\right\},
    \end{split}
\end{equation}
where we introduced the superoperators
\begin{equation}
    \mathscr{H}_\text{eff}\rho = \hat{H}_\text{eff}\rho + \rho\hat{H}_\text{eff}^\dagger,
    \quad
    \hat{H}_\text{eff} = \hat{H}_\text{LS} - \frac{i}{2}\sum_q\gamma_q\hat{\ell}_q^\dagger\hat{\ell}_q.
\end{equation}
Therefore, when the Lamb shift is \textit{not} neglected, the quantum Fisher information, and consequently the limiting activity as well, has the additional contribution given in Eq.~\eqref{app:eq:LS-QFI-extra}.

\section{Total dynamical activity in driven systems}\label{app:total-DA}
In this Appendix, we discuss the total dynamical activity in a driven system using the example of a two-level system with a time-dependent energy splitting driven at speed $v$ with Hamiltonian
\begin{equation}
    \hat{H}(t,v)= \hbar \omega \lambda(t v) \hat{\sigma}_z =\hbar \omega \lambda(t v) (\ket{e}\bra{e}-\ket{g}\bra{g} ).
\end{equation}
We consider a pure initial state $\ket{+} = (\ket{e} +\ket{g})/\sqrt{2}$ and a rescaling of time $\ket{\psi(\theta t,v)} = \hat{U}(\theta t,0;v) \ket{+}$ with the corresponding QFI
\begin{equation}\label{eq: TLS QFI time}
    \mathcal{F}_{\theta\theta}(t)\big|_{\theta =1} = t^2 \mathcal{F}_{tt}(t)  =4 t^2 \omega^2 \lambda^2(t v).
\end{equation}
This QFI is used to define a quantum activity in Refs.~\cite{Hasegawa2023May,Nakajima2023Nov, Nishiyama2024Apr,Yunoki2026Apr}. It should, however, be noted that basing a definition of activity on this parametrization and on QFI results in a quantity that is fully determined by the geometric speed at the final time and by the elapsed time. It does not probe the full driven evolution, and can be made to decrease over time if $\lambda(tv)$ decreases faster than $t$ grows~\cite{Palmqvist2026}. In particular, if $\lambda(tv)$ goes to zero, Eq.~\eqref{eq: TLS QFI time} is zero regardless of the earlier dynamics. 

For the total dynamical activity introduced in Sec. \ref{sec:relate_total}, the situation is different. From the rescaling $\ket{\psi(\theta t,v/\theta)}$ we find  
\begin{equation}\label{eq: TLS TDA}
\begin{aligned}
        \mathscr{A}(t,v) &=   \int_0^t \!\!\frac{ds}{\hbar}\int_0^t \!\!\frac{ds'}{\hbar}\langle \delta\hat{H}^\MRH(s,v)\delta\hat{H}^\MRH(s',v)\rangle \\
        & = \left(\int_0^t ds \omega \lambda(s v)\right)^2.
\end{aligned}
\end{equation}
Here $\mathscr{A}(t,v) $ depends on the entire history of the evolution. However, Eq.~\eqref{eq: TLS TDA} is not guaranteed to increase monotonically with time. In particular, if the drive is such that $\lambda(t v)$ changes sign during the evolution, it is even possible to find $\mathscr{A}(t_2,v) =0$ despite $\mathscr{A}(t_1,v) \neq 0$ for $t_1\leq t_2$. The mechanism for this nonmonotonicity is, however, completely different from the previous time-based definition. It occurs when the driving is such that the evolution is coherently reversed. This stems from the fact that the total dynamical activity is defined from the final state, and without introducing intermediate measurements during the evolution, there is no physically conceivable way to extract this information if the driving is such that the evolution is coherently reversed.

Moreover, a definition of activity based on outcomes of measurements performed on a driven quantum system can display nonmonotonicity if the final measurement time is variable. However, a measurement-based definition of activity will increase monotonically with the number of measurements performed since information is irreversibly transferred to a classical record, meaning that it cannot be lost due to later coherent evolution. For the PDA, this accumulating character of the classical dynamical activity is regained when the system is coupled to infinite reservoirs.

\section{\SKUR\  in DQD:   additional details and plots}\label{app: SKUR energy DQD}

\subsection{Particle transport}

\begin{figure}
    \centering
   \includegraphics{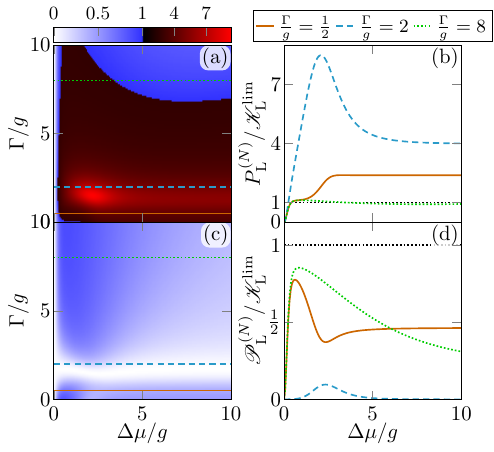}
\caption{(a,b) Violation of classical local KUR [Eq.~\eqref{eq:classical-KUR}] and (c,d) tightness of \SKUR\ [Eq.~\eqref{eq: ss transport relax MKUR}] in DQD for varying coupling strength $\Gamma = \Gamma_\MRL = \Gamma_\MRR$ and chemical potential bias $\Delta \mu$. We set $T_\MRL =T_\MRR =0.1 g/\kB$, and $\epsilon_d=0$.}
    \label{fig: DQD Gamma mu}
\end{figure}

We here present additional plots, complementing Fig.~\ref{fig: DQD g mu} and \ref{fig: DQD temperature} of the main text. A similar behavior to Fig.~\ref{fig: DQD g mu} is shown in Fig.~\ref{fig: DQD Gamma mu}, where we vary the bias $\Delta\mu$ and the symmetric coupling strength $\Gamma$.
Again, the classical local KUR is often violated, with a maximum at $2g\approx\Gamma$. The \SKUR\ is instead always valid, and approaches 1 for large values of $\Gamma$.

\subsection{Energy transport}

\begin{figure}[t]
    \centering
   \includegraphics{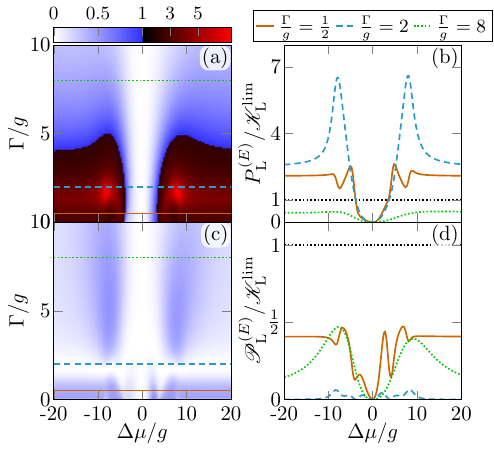}
\caption{
(a,b) Classical local KUR [Eq.~\eqref{eq:classical-KUR}] and (c,d) \SKUR\ [Eq.~\eqref{eq: ss transport relax MKUR}] for energy current in DQD for varying coupling strength $\Gamma = \Gamma_\MRL = \Gamma_\MRR$ and chemical potential bias $\Delta \mu= \mu_\MRL-\mu_\MRR$. We set $T_\MRL =T_\MRR =0.1 g/\kB$, and $\epsilon_d=3g$.
}
    \label{fig: DQD energy Gamma mu}
\end{figure}

\begin{figure}[t]
    \centering
   \includegraphics{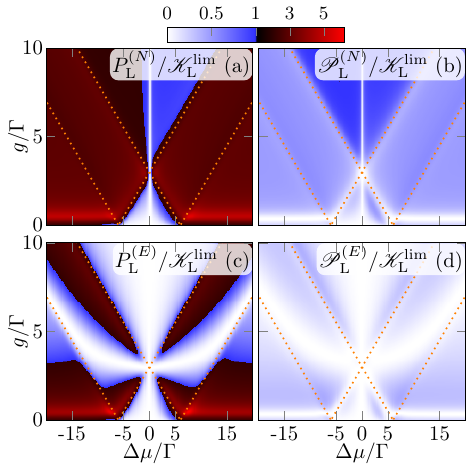}
\caption{
(a,c) Classical local KUR [Eq.~\eqref{eq:classical-KUR}] and (b,d) \SKUR\ [Eq.~\eqref{eq: ss transport relax MKUR}] for particle (a,b) and energy (c,d) currents in DQD for varying tunneling strength $g$ and chemical potential bias $\Delta \mu$.
We set $\Gamma_\MRL=\Gamma_\MRR =\Gamma$, $T_\MRL =T_\MRR =0.1 \Gamma/\kB$, and $\epsilon_d=3\Gamma$. The orange dotted lines lie at $g=\pm \frac{\Delta\mu}{2}-\epsilon_d$ and $g=\pm \frac{\Delta\mu}{2}+\epsilon_d$.
}
    \label{fig: app DQD energy current}
\end{figure}

Here, we show explicit expressions and additional plots for the \SKUR\ applied to the steady-state energy current into the left reservoir, complementing the results presented in the main paper. The parametrized energy current reads~\cite{Meir1992Apr}
\begin{equation}
    \begin{aligned}
        J^{(E)}_\mathrm{L}(\theta) &= \lim_{t\rightarrow\infty} \partial_t \langle \hat{H}^\MRH_\MRL(t,\theta) \rangle\\
       &= \frac{1}{\hbar} \int^\infty_{-\infty}  \frac{d \epsilon}{2 \pi} \epsilon \mathcal{T}_{\MRL\MRR}(\epsilon,\theta) (f_\MRR (\epsilon) -f_\MRL(\epsilon)).
    \end{aligned}
\end{equation}
The average current at $\theta=1$ and the susceptibility, entering the energy response current $\mathcal{J}^{(E)}_\MRL = \frac{1}{2}(J^{(E)}_\MRL +\dot{M}_\MRL^{(E)})$, are 
\begin{equation}
    J^{(E)}_\MRL = J^{(E)}_\MRL(\theta)\big|_{\theta=1}\quad \dot{M}_\MRL^{(E)} = \partial_\theta ( J^{(E)}_\MRL(\theta) /\theta )\big|_{\theta=1}.
\end{equation}
The zero-frequency noise in the energy current is expressed as~\cite{blanter_shot_2000}
\begin{equation}
\begin{aligned}
        S^{(E)}_\mathrm{L}&= \lim_{t\rightarrow\infty} \partial_t \mathrm{Var}[\hat{H}_\MRL^\mathrm{H}(t)-\hat{H}_\MRL^\mathrm{H}(0)]\\
    &=\frac{1}{\hbar} \int^{\infty}_{-\infty}\frac{d\epsilon}{2\pi} \epsilon^2\left\{  \mathcal{T}_{\MRL\MRR}(\epsilon) (F_{\MRL\MRR}(\epsilon) +F_{\MRR\MRL}(\epsilon) )\right\} \\
    &\quad-\frac{1}{\hbar} \int^{\infty}_{-\infty}\frac{d\epsilon}{2\pi} \epsilon^2\left\{  \mathcal{T}_{\MRL\MRR}(\epsilon) (f_{\MRR}(\epsilon) -f_{\MRL}(\epsilon) )\right\}^2.
\end{aligned}
\end{equation}

In Fig.~\ref{fig: DQD energy Gamma mu}(a,b), we show that the classical local KUR $(J^{(E)}_\mathrm{L})^2/S^{(E)}_\mathrm{L}  \leq \mathscr{K}^\mathrm{lim}_\MRL$ is violated, especially when $\Gamma\approx 2g$, just as for the particle current.
The \SKUR~\eqref{eq: ss transport relax MKUR} for the energy current is not as tight as the one for the particle current, see Fig.~\ref{fig: DQD energy Gamma mu}(c,d).
The comparison between particle current and energy current is highlighted in Fig.~\ref{fig: app DQD energy current}, where the precisions for the particle- [Fig.~\ref{fig: app DQD energy current}(a,b)] and energy-currents [Fig.~\ref{fig: app DQD energy current}(c,d)] are shown for the same choice of parameters.
The position of the dot's single-particle energy level with respect to the transport window $\Delta\mu$ is indicated by the dotted orange lines, and matches the features of both classical KUR and \SKUR.
Overall, these plots show that energy transport does not violate the classical KUR in as many parameter regions and not as strongly as particle transport. Furthermore, the \SKUR\ is not as tight for energy transport as it is for charge transport.

\begin{figure*}[ht!]
    \centering
    \includegraphics[width=6.6in]{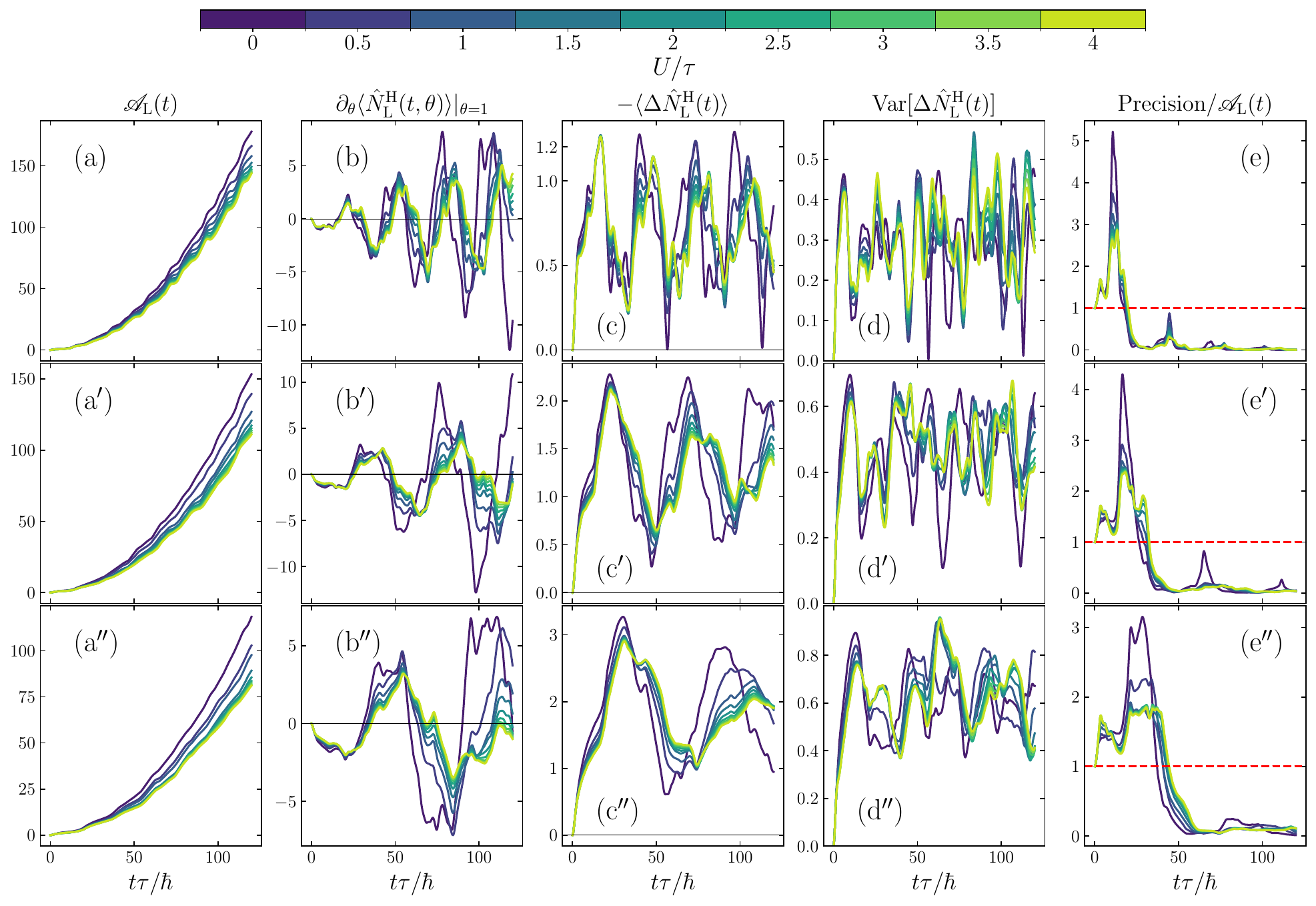}
    \caption{PDA $\mathscr{A}_\MRL(t)$, response in particle number, the negative of the mean transfers and its variance, and precision ${\langle \Delta\hat{N}^\mathrm{H}_\mathrm{L}(t)\rangle^2}/{ \mathrm{Var}[\Delta \hat{N}^\mathrm{H}_\mathrm{L}(t)]}$ divided by the PDA, showing that the quantum correction $\Phi^{(N)}_\MRL(t)$ plays a crucial role in the validity of the \SKUR. The first, second and third rows show leads with $W=3,6,9$ sites respectively and different color lines show different strengths of the Coulomb interactions $U$. All panels use $\tau_\MRL=\tau_\MRR=\tau/2$, $g=\tau$ and $\epsilon_d =0$.} 
    \label{fig:app: SKUR finite size}
\end{figure*}

\section{PDA at thermal equilibrium}\label{app:thermal}

Here, we calculate the PDA for a state in thermal equilibrium.
We consider a system at inverse temperature
$\beta=1/\kB T$ and the Hamiltonian $\hat{H}=\hat{H}_0 +\hat{V}= \sum_i \epsilon_i \ket{\epsilon_i}\bra{\epsilon_i}$ and express the density matrix as
\begin{equation}
    \hat{\rho} = \sum_i p_i \ket{\epsilon_i} \bra{\epsilon_i} ,\quad p_i = e^{-\epsilon_i \beta}/Z,\quad Z=\sum_i e^{-\epsilon_i \beta}.
\end{equation}
By introducing $\omega_{ij}=(\epsilon_i-\epsilon_j)/\hbar$ and using the Boltzmann factors~\cite{Liu2019Dec} we expand the integrands inside the square modulus and obtain
\begin{equation}
\begin{aligned}
          &\sum_{ij} \frac{(p_i -p_j)^2}{p_i+p_j} \bra{\epsilon_i} \hat{V}^\mathrm{H} (t) \ket{\epsilon_j} \bra{\epsilon_j} \hat{V}^\mathrm{H} (t') \ket{\epsilon_i}=\\
          &= \sum_{ij} \frac{ \tanh^2[\hbar\omega_{ij}\beta/2]}{Z} \left(e^{-\epsilon_i\beta} +e^{-\epsilon_j\beta}\right) V_{ij}V_{ji} e^{i\omega_{ij} (t-t')} ,
\end{aligned}
\end{equation}
where we defined $V_{ij} = \bra{\epsilon_i}\hat{V}\ket{\epsilon_j}$. Moreover,
\begin{equation}
  \begin{aligned}
        R_{VV}(t,t')&:=\langle \{ \hat{V}^\mathrm{H}(t),\hat{V}^\mathrm{H}(t') \}\rangle \\&= \sum_{ij} \frac{e^{-\epsilon_i\beta} +e^{-\epsilon_j\beta}}{Z}  V_{ij}V_{ji} e^{i\omega_{ij} (t-t')} .
  \end{aligned}
\end{equation}
Now, using $2 \pi \delta(\omega +\omega_{ij})= \int^\infty_{-\infty} dt e^{i(\omega + \omega_{ij} )t}$, we find
\begin{equation}
\begin{aligned}
        &R_{VV}(\omega,\omega') = \!\!\int_{-\infty}^{\infty}\!\! dt \int_{-\infty}^{\infty}\!\! dt' \langle \{\hat{V}^\mathrm{H}(t),\hat{V}^\mathrm{H}(t')  \} \rangle e^{- i \omega t- i \omega' t'}\\
         &= 4 \pi^2 \sum_{ij} \frac{e^{-\epsilon_i \beta} +e^{-\epsilon_j \beta}}{Z} V_{ij} V_{ji} \delta(\omega -\omega_{ij}) \delta(\omega' +\omega_{ij}).
\end{aligned}
\end{equation}
Since $[\hat{\rho},\hat{H}]=0$, the matrix elements of $\hat{V}^\MRH(s)$ in the eigenbasis of $\hat{\rho}$ are $V_{ij}e^{i\omega_{ij}s}$, such that
\begin{equation}
    \Big|\braket{\epsilon_i|\int_0^t\tfrac{ds}{\hbar}\hat{V}^\MRH(s)|\epsilon_j}\Big|^2=\frac{|V_{ij}|^2}{\hbar^2}\frac{4\sin^2(\omega_{ij}t/2)}{\omega_{ij}^2}.
\end{equation}
We introduce the symmetrized noise spectrum
\begin{equation}
\begin{aligned}
    S_{VV}(\omega)&=\int_{-\infty}^{\infty}d\tau\,e^{i\omega\tau}\tfrac12\langle\{\delta\hat{V}^\MRH(\tau),\delta\hat{V}^\MRH(0)\}\rangle\\
    &=2\pi\sum_{ij}\frac{p_i+p_j}{2}|\delta V_{ij}|^2\,\delta(\omega+\omega_{ij}),
\end{aligned}
\end{equation}
which is related to the expression above by $R_{VV}(\omega,\omega')=4\pi\,\delta(\omega+\omega')\,S_{VV}(\omega)$, up to the contribution of $\langle\hat{V}\rangle$ at $\omega=0$. With this, Eqs.~\eqref{eq:A_V} and~\eqref{eq:Alim} become
\begin{equation}
\begin{aligned}
    \mathscr{A}_V(t)&=\frac{1}{\hbar^2}\int\frac{d\omega}{2\pi}\,S_{VV}(\omega)\tanh^2\!\Big(\frac{\beta\hbar\omega}{2}\Big)\frac{4\sin^2(\omega t/2)}{\omega^2},\\
    \mathscr{A}^\mathrm{lim}_V(t)&=\frac{1}{\hbar^2}\int\frac{d\omega}{2\pi}\,S_{VV}(\omega)\,\frac{4\sin^2(\omega t/2)}{\omega^2}.
\end{aligned}
\end{equation}
For a continuous spectrum and $S_{VV}(\omega)$ regular at $\omega=0$, we use $4\sin^2(\omega t/2)/\omega^2\to2\pi t\,\delta(\omega)$ at long times, and find the rates
\begin{equation}
    \mathscr{K}_V=0,\qquad \mathscr{K}^\mathrm{lim}_V=\frac{S_{VV}(0)}{\hbar^2}.
\end{equation}
The factor $\tanh^2(\beta\hbar\omega/2)$ removes the low-frequency equilibrium fluctuations, $|\omega|\lesssim\kB T/\hbar$. These fluctuations contribute to $\mathscr{A}^\mathrm{lim}_V$ but carry no information on the coupling strength. In global equilibrium, the PDA rate hence vanishes, while the limiting activity rate is set by the zero-frequency noise of the coupling Hamiltonian.

\section{S-KUR in finite-size interacting system}\label{app: SKUR finite size}

\begin{figure}[t!]
    \centering
   \includegraphics[width=3.3in]{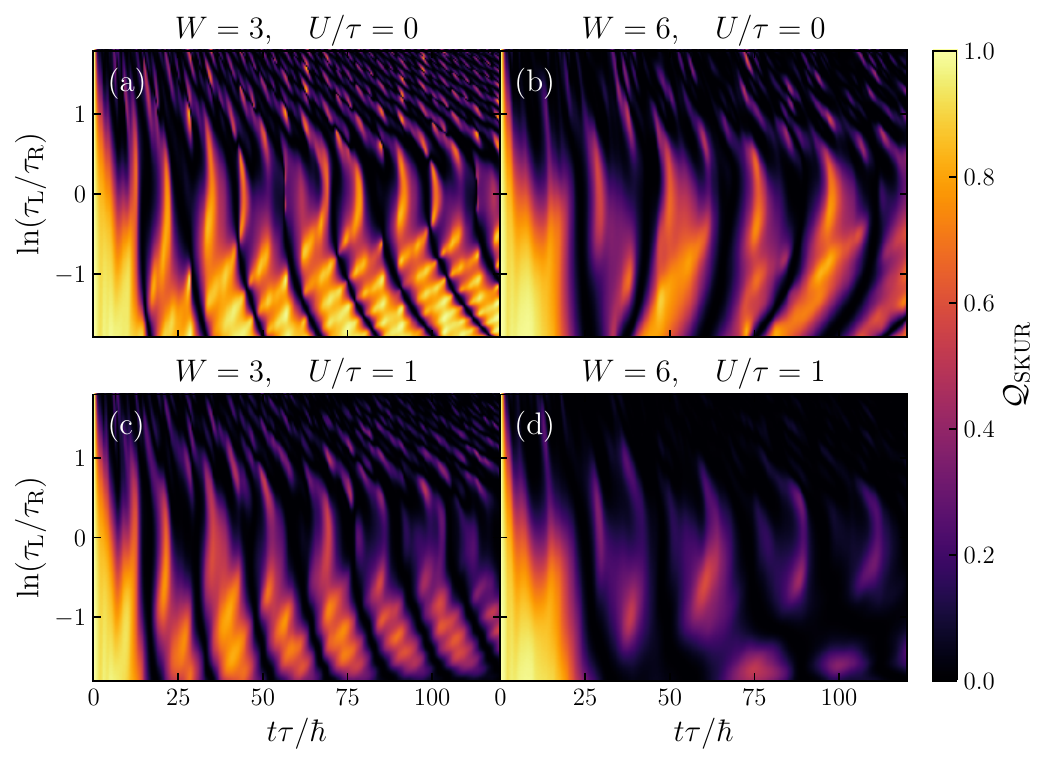}
    \caption{\SKUR\ ratio $\mathcal{Q}_\mathrm{SKUR}$ of Eq.~\eqref{eq: finite time SKUR ratio} as function of time and lead coupling asymmetry $\tau_\MRL/\tau_\MRR$ for varying value of Coulomb interaction strength and lead length. The lead couplings are parametrized by $\tau_\MRL =e^{v/2}\tau$, $\tau_\MRR =e^{-v/2} \tau$ with $v$ real. For all panels we use $g=\tau$ and $\epsilon_d =0$.}  
    \label{fig:app: SKUR finite size asymmetry}
\end{figure}

{In this Appendix, we provide additional details on the example of the \SKUR\ applied to the finite size system discussed in Sec.~\ref{sec: DQD finite leads}. 
In Fig.~\ref{fig:app: SKUR finite size} we display the PDA $\mathscr{A}_\text{L}(t)$, response of particle number $\partial_\theta\langle\hat{N}_\text{L}^\text{H}(t,\theta)\rangle|_{\theta=1}$, the negative of the average transfer out of L $(-\langle \Delta \hat{N}_\text{L}^\text{H}(t)\rangle)$, its variance $\text{Var}[\Delta \hat{N}_\text{L}^\text{H}(t)]$ and the ratio between the precision ${\langle \Delta\hat{N}^\mathrm{H}_\mathrm{L}(t)\rangle^2}/{ \mathrm{Var}[\Delta \hat{N}^\mathrm{H}_\mathrm{L}(t)]}$ and the PDA. Notice that the original formulation of the finite time KUR uses a total activity and the differential operator $t\partial_t$~\cite{DiTerlizzi2018Dec}, hence the results of panels~(e,e$'$,e$''$) should not be interpreted as probing the classical, finite-time KUR. 

The PDA increases over time for all displayed cases, meaning that the system becomes more sensitive to the coupling strength as it evolves.
Instead, the mean change and the response displays oscillations, as particles move back and forth along the chain.
It is however interesting to note that $\langle \Delta\hat{N}^\mathrm{H}_\mathrm{L}(t) \rangle$ dampens in time whereas $\partial_\theta \langle \Delta\hat{N}^\mathrm{H}_\mathrm{L}(t,\theta)\rangle|_{\theta=1}$ increase in magnitude over time. Moreover, in panels~(e,e$'$,e$''$), the ratio between precision and PDA takes on values above $1$, indicating that the susceptibility $M^{(N)}_\MRL(t)$ or, equivalently, the quantum correction $\Phi^{(N)}_\MRL(t)$, plays a key role in guaranteeing the validity of the \SKUR.
However, it is interesting to note that, for longer times, the $\mathcal{Q}_\mathrm{SKUR}$ in Fig.~\ref{fig: finite size SKUR}(a,a$'$,a$''$) reaches higher values than the ratio in~panels(e,e$'$,e$''$) of Fig.~\ref{fig:app: SKUR finite size}. This means that not only does the susceptibility term guarantee the general validity of the \SKUR\, it can also result in a tighter bound which traces back to the increasing magnitude of the responses oscillations in panels~(c,c$'$,c$''$) of Fig.~\ref{fig:app: SKUR finite size}.

In Fig.~\ref{fig:app: SKUR finite size asymmetry} we display the \SKUR\ ratio as function of coupling asymmetry and time, for different lead lengths $W$ and Coulomb interaction strengths $U$. The coupling asymmetry is parametrized as $\tau_\MRL = e^{v/2}\tau$ and  $\tau_\MRR = e^{-v/2}\tau$ with $v$ real. 
The parametrization keeps $\tau_\MRL\tau_\MRR=\tau^2$ fixed, while $\ln(\tau_\MRL/\tau_\MRR)=v$.
Three features stand out: First, the revivals of $\mathcal{Q}_\mathrm{SKUR}$ in time persist for all coupling asymmetries, with revival times that shift with $v$.
Second, $\mathcal{Q}_\mathrm{SKUR}$ reaches its highest values, close to $1$, even at long times, when the left lead is weakly coupled, $\tau_\MRL<\tau_\MRR$. This mirrors the steady-state results of Fig.~\ref{fig:SKUR sample}, where the \SKUR\ is tightest for $\Gamma_\MRL/\Gamma_\MRR\ll1$. For a weak link, the coupling strength enters the dynamics mainly through the particle transfers across this link, which are directly recorded by $\hat{N}_\MRL$. For strong coupling, $\tau_\MRL$ induces significant hybridization between the left lead and the DQD, which is not captured by the particle number.
Third, the Coulomb interaction damps the revivals, most strongly for the longer leads [Fig.~\ref{fig:app: SKUR finite size asymmetry}(d)]. For $U\neq0$, the many-body energies are no longer sums of single-particle energies, and, consequently, the coherent oscillations underlying the revivals dephase.}

\bibliography{refs.bib}

\end{document}